\documentclass[10pt,conference]{IEEEtran}
\usepackage{cite}
\usepackage{amsmath,amssymb,amsfonts}
\usepackage{graphicx}
\usepackage{textcomp}
\usepackage{xcolor}
\usepackage[hyphens]{url}
\usepackage{fancyhdr}
\usepackage{hyperref}

\newcommand{\hpcayear}{2025}

\newcommand{\zy}{\textcolor{black}}

\newcommand{\cready}{\textcolor{black}}

\usepackage{algorithm}
\usepackage{algpseudocode}
\usepackage{booktabs}
\usepackage{multirow}

\usepackage{pifont}
\usepackage{listings}
\usepackage{xcolor}

\definecolor{codegreen}{rgb}{0,0.6,0}
\definecolor{codegray}{rgb}{0.5,0.5,0.5}
\definecolor{codepurple}{rgb}{0.58,0,0.82}
\definecolor{backcolour}{rgb}{0.95,0.95,0.92}

\lstdefinestyle{mystyle}{
    basicstyle=\ttfamily\footnotesize,
    breakatwhitespace=false,         
    breaklines=true,                 
    captionpos=b,                    
    keepspaces=true,                 
    numbersep=5pt,                  
    showspaces=false,                
    showstringspaces=false,
    showtabs=false,                  
    tabsize=2
}

\usepackage[textsize=footnotesize,textwidth=1.05\linewidth]{todonotes}
\makeatletter
\if@todonotes@disabled

\else

\fi
\makeatother

\newcommand{\hpcasubmissionnumber}{1671}
\title{Gaussian Blending Unit: An Edge GPU Plug-in for Real-Time Gaussian-Based Rendering in AR/VR}
\def\hpcacameraready{} % Uncomment to build camera-ready version

\newcommand\hpcaauthors{Zhifan Ye, Yonggan Fu, Jingqun Zhang, Leshu Li, Yongan Zhang, Sixu Li, \\ Cheng Wan, Chenxi Wan, Chaojian Li, Sreemanth Prathipati, and Yingyan (Celine) Lin}
\newcommand\hpcaaffiliation{Georgia Institute of Technology}
\newcommand\hpcaemail{\{zye327, yfu314, jzhang3368, yzhang919, sli941, \\ cwan39, cli851, sreemanth, celine.lin\}@gatech.edu}

\author{
  \ifdefined\hpcacameraready
    \IEEEauthorblockN{\hpcaauthors{}}
      \IEEEauthorblockA{
        \hpcaaffiliation{} \\
        \hpcaemail{}
      }
  \else
    \IEEEauthorblockN{\normalsize{HPCA \hpcayear{} Submission
      \textbf{\#\hpcasubmissionnumber{}}} \\
      \IEEEauthorblockA{
        Confidential Draft \\
        Do NOT Distribute!!
      }
    }
  \fi 
}

\fancypagestyle{camerareadyfirstpage}{%
  \fancyhead{}
  
  \fancyhead[C]{
    \ifdefined\aeopen
    \parbox[][12mm][t]{13.5cm}{\hpcayear{} IEEE International Symposium on High-Performance Computer Architecture (HPCA)}    
    \else
      \ifdefined\aereviewed
      \parbox[][12mm][t]{13.5cm}{\hpcayear{} IEEE International Symposium on High-Performance Computer Architecture (HPCA)}
      \else
      \ifdefined\aereproduced
      \parbox[][12mm][t]{13.5cm}{\hpcayear{} IEEE International Symposium on High-Performance Computer Architecture (HPCA)}
      \else
      \parbox[][0mm][t]{13.5cm}{\hpcayear{} IEEE International Symposium on High-Performance Computer Architecture (HPCA)}
    \fi 
    \fi 
    \fi 
    \ifdefined\aeopen 
      \includegraphics[width=12mm,height=12mm]{ae-badges/open-research-objects.pdf}
    \fi 
    \ifdefined\aereviewed
      \includegraphics[width=12mm,height=12mm]{ae-badges/research-objects-reviewed.pdf}
    \fi 
    \ifdefined\aereproduced
      \includegraphics[width=12mm,height=12mm]{ae-badges/results-reproduced.pdf}
    \fi
  }
  \fancyfoot[C]{}
}
\begin{document}
\maketitle

%Enables the camera ready header and footer
\ifdefined\hpcacameraready 
  \thispagestyle{camerareadyfirstpage}
  \pagestyle{empty}
\else
  \thispagestyle{plain}
  \pagestyle{plain}
\fi

\newcommand{\hpcaheight}{0mm}
\ifdefined\eaopen
\renewcommand{\hpcaheight}{12mm}
\fi

%%%%%%%%%%%%%%%%%%%%%%%%%%%%%%%%%%%%%%%%
%%%%%%%% -- PAPER CONTENT STARTS -- %%%%%%%%%

\begin{abstract}

The rapidly advancing field of Augmented and Virtual Reality (AR/VR) demands real-time, photorealistic rendering on resource-constrained platforms. 3D Gaussian Splatting, delivering state-of-the-art (SOTA) performance in rendering efficiency and quality, has emerged as a promising solution across a broad spectrum of AR/VR applications. However, despite its effectiveness on high-end GPUs, it struggles on edge systems like the Jetson Orin NX Edge GPU, achieving only 7-17 FPS—well below the over 60 FPS standard required for truly immersive AR/VR experiences. Addressing this challenge, we perform a comprehensive analysis of Gaussian-based AR/VR applications and identify the Gaussian Blending Stage, which intensively calculates each Gaussian's contribution at every pixel, as the primary bottleneck. In response, we propose a Gaussian Blending Unit (GBU), an edge GPU plug-in module for real-time rendering in AR/VR applications. Notably, our GBU can be seamlessly integrated into conventional edge GPUs and collaboratively supports a wide range of AR/VR applications. Specifically, GBU incorporates an intra-row sequential shading (IRSS) dataflow that shades each row of pixels sequentially from left to right, utilizing a two-step coordinate transformation. This transformation enables (1) the sharing of intermediate values between adjacent pixels, reducing pixel-wise computation costs by up to 5.5$\times$, and (2) the early identification and skipping of Gaussians that minimally contribute to the pixels, reducing per-pixel computation by up to 93\%. When directly deployed on a GPU, the proposed dataflow achieved a non-trivial 1.72$\times$ speedup on real-world static scenes, though still falls short of real-time rendering performance. Recognizing the limited compute utilization in the GPU-based implementation, GBU enhances rendering speed with a dedicated rendering engine that balances the workload across rows by aggregating computations from multiple Gaussians. Additionally, GBU integrates a Gaussian Reuse Cache, reducing off-chip memory accesses by 44.9\% and resulting in a 1.14$\times$ speedup in rendering. Experiments across representative AR/VR applications demonstrate that our GBU provides a unified solution for on-device real-time rendering while maintaining SOTA rendering quality.

% Addressing this challenge, we perform a comprehensive analysis of Gaussian-based AR/VR applications and identify the Gaussian Blending Stage, which intensively calculates each Gaussian's contribution at every pixel, as the primary bottleneck. In response, we have developed the Gaussian Blending Unit (GBU), an edge GPU plug-in module for real-time rendering in AR/VR applications. Notably, the GBU can be seamlessly integrated into conventional edge GPUs and collaboratively supports a wide range of AR/VR applications. \zy{We leverage a key insight: through appropriate coordinate transformation, we can (1) efficiently filter out Gaussians that contribute marginally to the pixel color and (2) leverage shared computation between adjacent pixels, both in a row-wise computation manner. Motivated by the insight, we propose a row-wise computation dataflow that reduces XX\% FLOPs through redundancy skipping and compute sharing. To fully unleash the benefits of this row-wise dataflow,} we develop a dedicated hardware module featuring: \zy{(1) asynchronized \textit{row-dispatching engines} and \textit{row-shading engines} that amortize the imbalanced workload between rows across multiple Gaussians; (2) a dedicated \textit{Gaussian reuse cache} that precisely caches the reusable Gaussians for reduced memory traffic.} Experiments across popular AR/VR applications demonstrate that our GBU provides a unified solution for on-device real-time rendering while maintaining SOTA rendering quality.

\end{abstract}

\section{Introduction}
\label{sec:intro}

The Augmented and Virtual Reality (AR/VR) sector is rapidly expanding, driven by substantial industry interest and investment~\cite{AppleVis7, MetaQues16}. Edge AR/VR platforms, like headsets, strive to provide immersive and interactive experiences in various applications such as virtual meetings, tourism, and try-ons. These applications require real-time, photorealistic rendering of scenes composed of static\cite{mildenhall2021nerf} and dynamic~\cite{Pumarola_2021_CVPR, luiten2023dynamic} objects, as well as human avatars with complex poses and expressions~\cite{holoportation, lombardi2018deep, Ma_2021_CVPR}. Therefore, it is crucial to develop a versatile rendering pipeline that can accurately reconstruct diverse real-world scenes and perform efficiently on edge AR/VR devices.

% \begin{figure}[!t]
% \centering
% \includegraphics[width=1.0\linewidth]{Figures/applications.pdf}
% \vspace{-2em}
% \caption{Examples of popular AR/VR applications~\cite{holoportation,Feng2022scarf}.}
% \label{fig:motivating_applications}
% \vspace{-1em}
% \end{figure}

\begin{figure}[!t]
\centering
\includegraphics[width=0.9\linewidth]{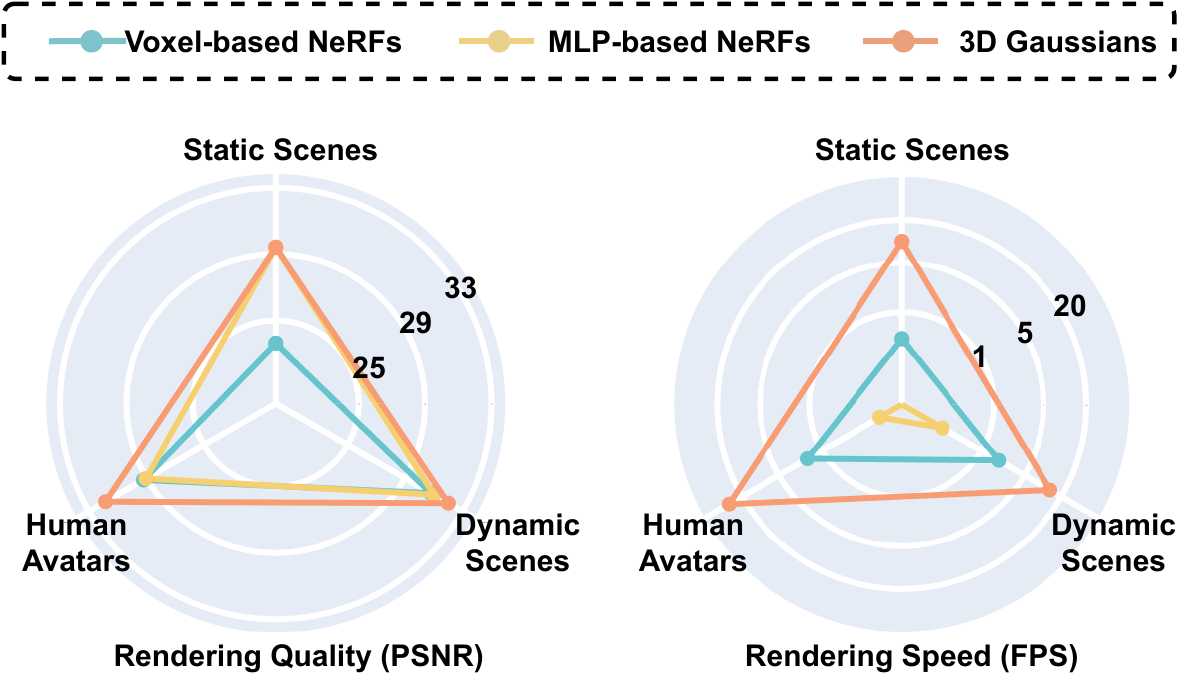}
\vspace{-1em}
\caption{Benchmarking 3D Gaussian Splatting~\cite{kerbl3Dgaussians, yang2023gs4d, SplattingAvatar} with previous works~\cite{mueller2022instant, barron2021mipnerf, wang2022mixed, attal2023hyperreel, jiang2022instantavatar, chen2021animatable} on real-world datasets~\cite{DeepBlending2018, li2022neural, chen2021animatable}. Here Peak-signal-to-noise-ratio (PSNR) and frames-per-second (FPS) are the reported ones from previous papers~\cite{kerbl3Dgaussians, yang2023gs4d, SplattingAvatar}, and all rendering speeds are measured on an edge GPU~\cite{JetsonOr73:online}.} 
\label{fig:motivating_benchmark}
\vspace{-1.5em}
\end{figure}

In the graphics and computer vision community, 3D Gaussian Splatting~\cite{kerbl3Dgaussians} has emerged as a highly promising 3D scene representation for AR/VR. It achieves SOTA reconstruction performance across various objects and scenes~\cite{kerbl3Dgaussians, wu20234dgaussians, li2023human101}, and excels in AR/VR tasks beyond 3D reconstruction, including 3D asset creation~\cite{tang2023dreamgaussian, zhang2024gaussiancube}, scene editing~\cite{chen2023gaussianeditor,huang2023sc}, and open vocabulary querying~\cite{qin2023langsplat}. Compared to previous representations like neural radiance fields (NeRFs)~\cite{mildenhall2021nerf}, 3D Gaussians offer better balance with faster reconstruction speed and significantly improve rendering framerate\cite{chen2024survey}, as shown in Fig.~\ref{fig:motivating_benchmark}. This makes 3D Gaussian Splatting an excellent option for on-device 3D applications on resource-constrained AR/VR platforms, where edge GPUs are the primary rendering hardware\cite{AppleVis7, MetaQues16}.

Despite the potential of 3D Gaussians for real-time rendering on server and desktop devices, a significant performance gap remains for real-time rendering (i.e., $\geq$ 60 FPS~\cite{framerate}) on edge devices. For example, rendering real-world scenes from the MipNeRF-360 dataset~\cite{barron2021mipnerf} on the Jetson Orin NX~\cite{JetsonOr73:online}, an edge GPU from NVIDIA, achieves only 7 to 17 FPS. This gap hinders the adoption of emerging AR/VR applications that leverage the latest in 3D reconstruction technology.
To bridge this gap, we conducted a comprehensive analysis of multiple 3D Gaussian-based rendering pipelines targeting various AR/VR applications~\cite{kerbl3Dgaussians, wu20234dgaussians, lei2023gart}. We identified that the Gaussian Blending stage, where the opacity of each Gaussian's contribution to pixels is calculated, is consistently the primary latency bottleneck, accounting for 48\% to 78\% of the rendering time in these applications. This stage requires intensive per-pixel processing, which involves multiple matrix-vector multiplications and becomes the bottleneck for overall latency.

To this end, we have developed the Gaussian Blending Unit (GBU), a hardware module designed for edge GPUs to facilitate real-time rendering using 3D Gaussians, enhancing AR/VR applications. This unit integrates smoothly with existing edge GPUs, accelerating the common rendering bottleneck to improve performance across various applications, with a design that ensures compatibility and scalability. Specifically, our contributions are summarized as follows:

\begin{itemize}

\item We conducted a comprehensive analysis of Gaussian-based AR/VR applications, characterizing their rendering pipelines into common and application-specific stages. Through this, we identified the Gaussian Blending stage as the common bottleneck that prohibits real-time rendering on resource-constrained AR/VR devices.

\item \zy{We developed GBU, a plug-in module for edge GPUs that enhances real-time rendering for AR/VR applications. By offloading the shared bottleneck, the Gaussian Blending stage, to GBU, we ensure real-time rendering speeds. Meanwhile, by keeping application-specific computations on the GPU's general-purpose compute units and seamless integration between GBU and GPU, the acceleration system maintains compatibility with a wide range of AR/VR applications.}

% 2024.07.29
% \item \zy{ In order to reduce the high computational volume of the bottleneck Gaussian Blending stage, GBU adopts a row-wise sequential computing dataflow dubbed \textit{IRSS (Intra-Row Sequential Shading)} dataflow. This dataflow is enhanced with coordinate transformations which allow sharing compute between adjacent pixels and efficiently skipping redundant computation.}

% 2024.07.30
\item \zy{To reduce the high computational load of the bottleneck Gaussian Blending stage, we developed an \textit{Intra-Row Sequential Shading (IRSS)} dataflow that sequentially shades each row of pixels from left to right. Enhanced with a two-step coordinate transformation, this dataflow reduces pixel-wise computational costs by up to 5.5$\times$ through the sharing of computations between adjacent pixels. Additionally, the IRSS dataflow allows the GBU to identify and skip redundant Gaussians that contribute minimally to a pixel, saving up to 92.3\% of the computational workload in the Gaussian Blending stage. When integrated with a GPU, the proposed IRSS dataflow increases rendering speed on real-world static scenes from 13 FPS to 22 FPS.}

% 2024.07.29
% \item \zy{ However, the proposed \textit{IRSS} dataflow only achieved limited speed up on GPUs, mainly due to (1) a low compute utilization under imbalanced workload between rows and (2) excessive random memory accesses for input Gaussian features. To fully unleash the efficiency benefits of the proposed dataflow and achieve real-time rendering, GBU equips (1) a row-centric tile engine that amortizes the unbalanced workload across multiple Gaussians and (2) a dedicated Gaussian reuse cache which features an optimized cache replacement policy guided by pre-computed reuse distance of Gaussian features.}

% 2024.07.30
\item \zy{To meet the real-time framerate requirements (i.e. over 60 FPS) of AR/VR applications~\cite{framerate}, we identified the imbalanced workload between pixel rows as a primary performance bottleneck in GPUs, resulting in only 18.9\% GPU utilization on real-world static scenes. To address this issue, the GBU hardware incorporates a dedicated rendering engine that mitigates the problem by enabling asynchronous rendering of individual rows and distributing the workload across multiple Gaussians. Additionally, the GBU is equipped with a specialized Gaussian Reuse Cache, which reduces off-chip memory accesses by 44.9\%, leading to a 1.14$\times$ speedup on real-world static scenes.}

\item We conducted extensive evaluations of our GBU across a variety of popular AR/VR rendering pipelines and applications, encompassing static and dynamic objects/scenes, as well as human avatars. The experiment results show that the GBU provides a comprehensive solution for real-time rendering, achieving speeds greater than 60 FPS on edge devices across a broad range of AR/VR applications, while consistently maintaining SOTA rendering quality.

\end{itemize}

It is worth noting that our \zy{\textit{IRSS} dataflow and }GBU design is applicable to a wide range of 3D applications on edge devices beyond just AR/VR platforms, thus inspiring future innovations in hardware and system support for Gaussian-based rendering and facilitating ubiquitous 3D intelligence on edge devices.

\section{Preliminaries of 3D Gaussians}

\label{sec:pre}

In this section, we first elaborate on reconstructing a static scene using 3D Gaussians and describe the corresponding rendering pipeline in Sec.~\ref{sec:pre:static} and Sec.~\ref{sec:pre:rendering}, respectively. We then detail how to extend the rendering pipeline to other popular AR/VR applications in Sec.~\ref{sec:pre:dynamic} and summarize their common workload patterns in Sec.~\ref{pre:pipeline}.

\subsection{Gaussian Splatting for Static Scene Reconstruction}
\label{sec:pre:static}

\begin{figure}[t]
\vspace{-0em}
\begin{center}
\includegraphics[width=\linewidth]{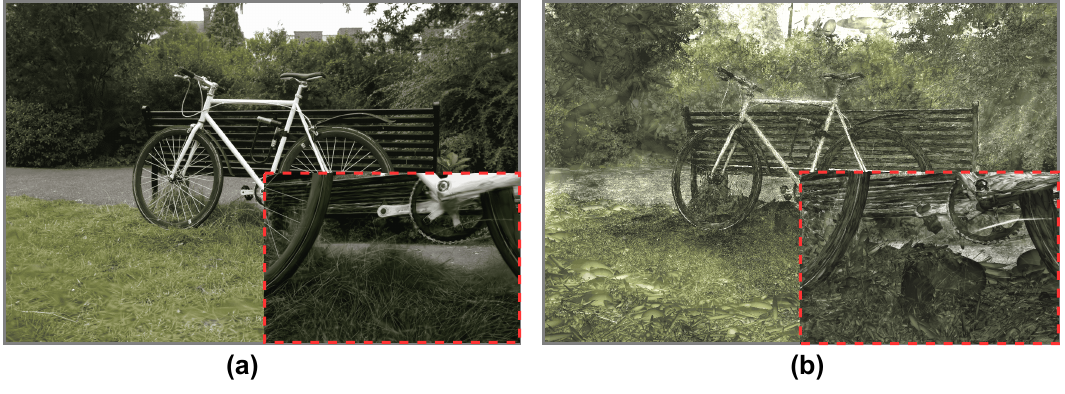}
 \vspace{-2em}
\caption{(a) A real-world image~\cite{barron2021mipnerf} rendered with 7 millions of 3D Gaussians; and (b) the corresponding 3D Gaussians. The red boxes highlight zoomed-in views for better visualization.}
\label{fig:gaussian_simple}
\end{center}
 \vspace{-2em}
\end{figure}

\begin{figure*}[!ht]
\centering
% \vspace{1em}
\includegraphics[width=1.0\linewidth]{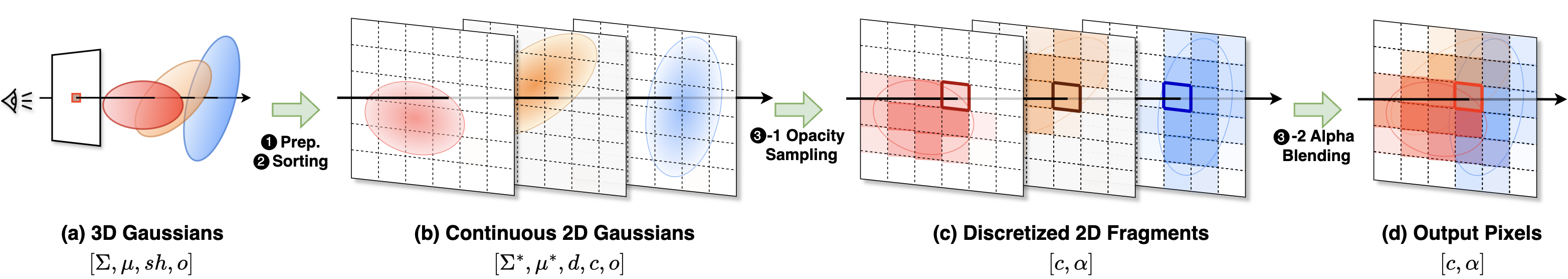}
\vspace{-2.0em}
% \caption{The rendering pipeline of 3D Gaussians.}
\caption{\zy{An illustration of the rendering pipeline for 3D Gaussian Splatting~\cite{kerbl3Dgaussians}. (a) A set of 3D Gaussians, each parameterized by a 3D Gaussian function (covariance $\Sigma$ and mean $\mu$), SH coefficients $sh$, and an opacity factor $o$. (b) Projected 2D Gaussians, each characterized by a 2D covariance $\Sigma^*$, a 2D mean $\mu^*$, depth $d$, color $c$, and opacity factor $o$. (c) Each 2D Gaussian is sampled at pixels, resulting in a set of 2D fragments with color $c$ and opacity $\alpha$. (d) Color $c$ and opacity $\alpha$ of output pixels are determined by accumulating the color and opacity of all fragments from all Gaussians that overlap each pixel.}}

%First, 3D Gaussian kernels are projected into continuous 2D kernels and sorted in depth order. Next, these 2D Gaussian kernels are sampled at each pixel center. Finally, the sampled color and opacity are blended at each pixel.}}
\label{fig:rendering_pipeline}
\vspace{-1.0em}
\end{figure*}

\noindent \textbf{Representing Scenes with 3D Gaussians.}
Recently, 3D Gaussian Splatting~\cite{kerbl3Dgaussians} has emerged as the SOTA method for 3D reconstruction tasks, excelling in both rendering quality and speed. As illustrated in Fig.~\ref{fig:gaussian_simple}, 3D Gaussian Splatting reconstructs 3D objects and scenes using a set of elliptical 3D Gaussian kernels, each described by an (unnormalized) Gaussian probability density function:
\begin{align}
\label{eq:gaussian_function}
G(x) = e^{-\frac{1}{2}(x-\mu)^T\Sigma^{-1}(x-\mu)},
\end{align}
where the Gaussian function is characterized by a 3D covariance matrix $\Sigma$ and is centered at the point $\mu$ (i.e., its mean). This covariance matrix can be decomposed into a rotation matrix $R$ and a scaling matrix $S$ via the decomposition $\Sigma=R^TS^TSR$, which effectively determines the orientation $R$ and scale $S$ of the Gaussian distribution in 3D space.

To represent the color and density distribution of a 3D object or scene, each Gaussian kernel is further assigned an opacity factor $o$ and a set of Spherical Harmonics (SH) coefficients $sh$. These SH coefficients $sh$ are utilized to determine the color $c=f(v; sh)$ of a Gaussian when viewed from a specific direction $v$, where $f$ is the spherical harmonics function~\cite{yu2022plenoxels}. Employing SH coefficients, rather than assigning a static, view-independent color to each Gaussian, enables the modeling of real-world visual phenomena, e.g., specular reflection and the Fresnel Effect. Collectively, tens of thousands of these Gaussian kernels represent colored 3D objects and scenes in complex real-world scenarios, as shown in Fig.~\ref{fig:gaussian_simple}(b).

\subsection{The Rendering Pipeline of 3D Gaussians}
\label{sec:pre:rendering}

Given reconstructed 3D Gaussians and a viewing direction, Fig.~\ref{fig:rendering_pipeline} illustrates the rendering pipeline, which transforms/renders the 3D Gaussian kernels into a 2D RGB image. This rendering process is divided into three steps:

\noindent\textbf{Rendering Step~\ding{182}: Preprocessing.} 
This preprocessing step serves two purposes: (1) projecting all 3D Gaussians onto the 2D screen and (2) computing the depth $d$ and RGB color $c$ of each Gaussian based on the view direction. After this step, each 3D Gaussian is transformed into a 2D Gaussian on the screen with an RGB color $c$ and a depth value $d$.
Similar to 3D Gaussians, each 2D Gaussian kernel is also defined by a covariance matrix $\Sigma^\star$ and centered at a point $\mu^\star$ (i.e., the mean value):
\begin{align}
\label{eq:gaussian_function_2d}
G^\star(x) = e^{-\frac{1}{2}(x-\mu^\star)^T{\Sigma^\star}^{-1}(x-\mu^\star)},
\end{align}
where its 2D covariance matrix $\Sigma^\star$ and mean value $\mu^\star$ can be derived from the corresponding 3D covariance $\Sigma$ and mean $\mu$ using the following formulas~\cite{zwicker2001ewa}:
\begin{align}
\label{eq:gaussian_projection}
\mu^\star = J W \mu;~~~\Sigma^\star = J W \Sigma W^T J^T.
\end{align} 
Here, $W$ represents a viewing transformation matrix that transforms Gaussians to the view space, and $J$ is a Jacobian matrix that defines the mapping from the 3D space to the 2D screen. A byproduct of the projection $W \mu$ is the depth $d$ of the Gaussian, i.e., the distance from the \cready{viewpoint} to the Gaussian's center. This depth is used to determine the occlusion relationships between Gaussians in subsequent steps. Concurrently, this step accounts for the viewing direction $v$ of the camera to compute the RGB color $c$ for all Gaussians, as specified in $c=f(v; sh)$.

\noindent\textbf{Rendering Step~\ding{183}: Sorting by Depth.} 
After projecting all the Gaussians onto the 2D screen, a pixel may overlap with multiple 2D Gaussians. Considering that Gaussians closer to the screen can occlude those farther away, the color and opacity of the overlapping 2D Gaussians should be blended based on a near-to-far depth order. Therefore, before the blending process in the next step, depth sorting is necessary to determine the blending order for the overlapping 2D Gaussians.

\noindent\textbf{Rendering Step~\ding{184}: Gaussian Blending.} 
This step blends the color and opacity of 2D Gaussians at each pixel according to a near-to-far depth order, producing the final 2D RGB image. As illustrated in Fig.~\ref{fig:rendering_pipeline}(c) and (d), this step involves two consecutive operations: \ding{184}-1 \textit{opacity computation}, which calculates the contribution, i.e., the opacity, of each Gaussian at the pixel; and \ding{184}-2 \textit{$\alpha$-blending}, which blends the colors of all Gaussians overlapping a pixel, weighted by their opacity at that pixel.

For \ding{184}-1 \textit{opacity computation}, each 2D Gaussian function is sampled at the pixel centers of the screen, resulting in a pixel-aligned 2D grid as shown in Fig.~\ref{fig:rendering_pipeline}(c). Each cell of this grid is referred to as a fragment (i.e., the footprint of a 2D Gaussian on a pixel, one pixel can have multiple fragments if multiple 2D Gaussians are projected onto it). Specifically, for a fragment centered at pixel $P$ and associated with Gaussian $G_i$, its color $c_{P,i}$ is the Gaussian's color $c_i$. Its opacity is determined by sampling the Gaussian function at pixel $P$, weighted by its opacity factor $o_i$:
\begin{align}
\alpha_{P,i} &= o_i G_i^\star(P) \\
&= o_i e^{-\frac{1}{2}(P-{\mu}^\star_i)^T{{\Sigma}^\star_i}^{-1}(P-{\mu}^\star_i)}.
\label{eq:alpha}
\end{align}
Subsequently, the \ding{184}-2 \textit{$\alpha$-blending} process blends the fragments overlapping the same pixel from all 2D Gaussians:
\begin{align}
\label{eq:gaussian_function}
\mathbb{C}_P = \sum_{i=1}^{n}T_{P,i}\alpha_{P,i}c_i,
\end{align}
where $i$ iterates over all fragments overlapping the same pixel $P$, ordered by depth, and $T_{P,i} = \prod_{j=1}^{i-1}(1-\alpha_{P,j})$ represents the accumulated transmittance, quantifying the occlusion effects caused by the first $(i-1)$ Gaussians.

\noindent\textbf{Practical Implementation of the Above Rendering Step \ding{184}.}
It is computationally infeasible to exhaustively compute and blend the contribution of each Gaussian to every pixel in the entire image. Therefore, in practice, the implementation of 3D Gaussian Splatting~\cite{kerbl3Dgaussians} truncates the 2D Gaussian function in Eq.~\ref{eq:gaussian_function_2d} at a predetermined threshold. Due to their negligible impact, fragments outside this truncation range (i.e. whose opacity is lower than the threshold) are not considered in $\alpha$-\textit{blending}. In other words, each 3D Gaussian kernel is truncated into a 2D ellipsoid when projected onto the 2D screen.

3D Gaussian Splatting~\cite{kerbl3Dgaussians} implemented a highly optimized CUDA kernel to blend these 2D ellipsoids. Specifically, it employs tile-based rendering, a strategy commonly used in mobile devices, which divides the screen into multiple $16 \times 16$ tiles and runs \ding{184}-1 and \ding{184}-2 in a per-tile basis. Each 2D ellipsoid is assigned to corresponding tiles based on its overlap with the tile. Each tile is then managed by a Streaming Multiprocessor (SM) on the GPU. The SM processes all the assigned Gaussians following the depth order, computes their contributions to all the pixels within the tile in parallel (using one thread per pixel), and updates the pixel colors.

\begin{figure}[!t]
\centering
\includegraphics[width=1.0\linewidth]{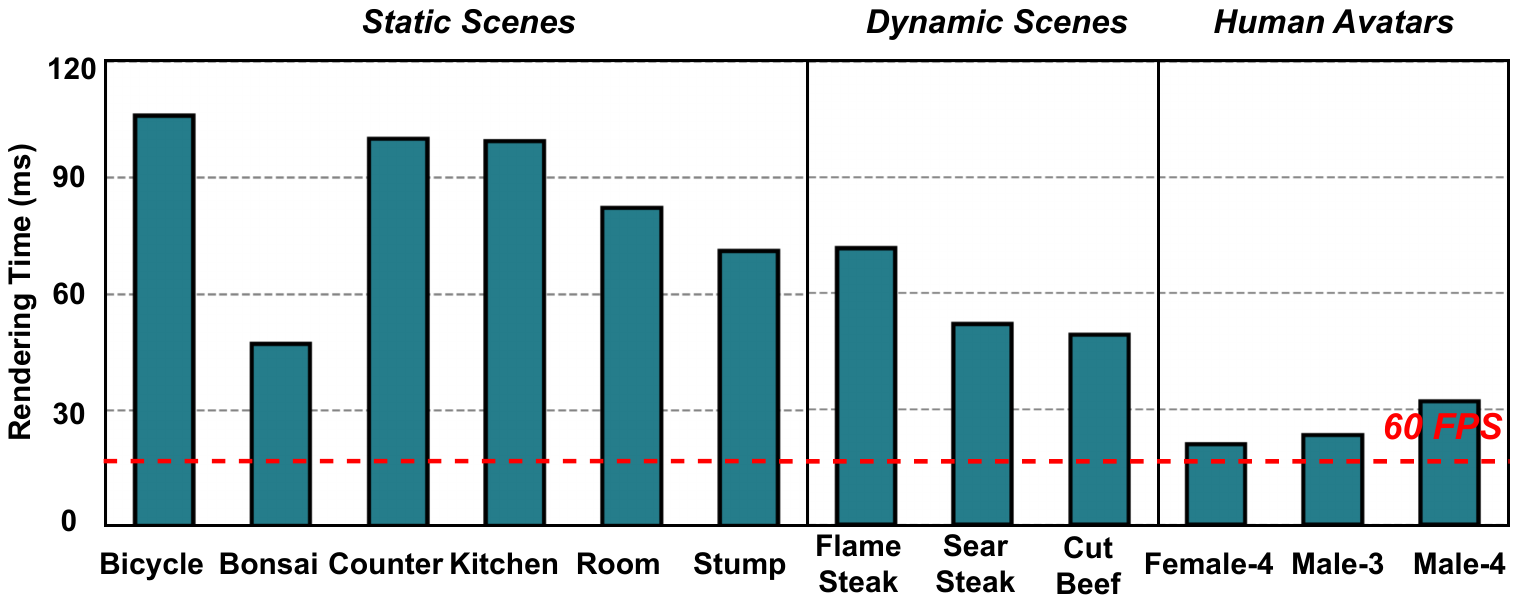}
\vspace{-2em}
\caption{End-to-end rendering time for three real-world datasets~\cite{barron2021mipnerf, Li_2022_CVPR, alldieck2018video}. The red line represents the maximum rendering time required to achieve real-time rendering (60 FPS).} 
\label{fig:profiling_breakdown_1}
\vspace{-1em}
\end{figure}

\subsection{Extend 3D Gaussians to More AR/VR Applications}
\label{sec:pre:dynamic}

In this subsection, we introduce the extensions of 3D Gaussians to more AR/VR applications, using dynamic scenes and avatars as examples. We will demonstrate the effectiveness of our techniques in these AR/VR applications in Sec.~\ref{sec:exp}.

\noindent\textbf{3D Gaussians for Dynamic Scene Reconstruction.} 
Besides static scenes, the 3D reconstruction of dynamic scenes is a highly desirable functionality in AR/VR applications involving evolving objects and scenes, such as remote education and virtual meetings. The complexity lies in capturing the intricate motion and deformation over time. 3D Gaussians, which explicitly decompose a scene into 3D Gaussian kernels, can effectively model both aspects. 
Specifically, to model the motion and deformation in dynamic scenes, it's crucial to make the Gaussian parameters time-dependent. For example, recent work on 4D Gaussian Splatting~\cite{yang2023gs4d} parameterizes a dynamic scene with a set of 4D Gaussian functions. Each 4D Gaussian kernel is defined by a 4D covariance matrix and a 4D mean value. The kernels can be efficiently sampled at any timestep \( t \), resulting in a set of 3D Gaussian kernels at the timestep. 
% This work also replaces the SH in 3D Gaussian Splatting with 4D Spherindrica Harmonics (4DSH) that takes both the time step $t$ and the viewing direction $v$ as input to determine the color of a Gaussian.

\noindent\textbf{3D Gaussians for Human Avatar Reconstruction.} 
Animating human avatars is another important task in AR/VR. Here, ``animatable" means that the reconstructed human avatar can be deformed according to given pose and expression parameters $\theta$, allowing for control over the avatar's pose and expression. Rendering human avatars at real-time framerates is crucial for many AR/VR applications, including telecommunications, virtual meetings, and remote education. A plethora of works~\cite{li2023human101, lei2023gart, zielonka2023drivable, hu2023gauhuman} has explored the application of 3D Gaussians to this domain. 
% The basic design methodology involves using \textit{forward skinning} to transform the 3D Gaussians based on the pose parameters \(\theta\), which define the transformations (rotation and translation) of human bones in that pose. After transformation, the 3D Gaussian kernels can represent the color and geometry of the human avatar at the specified pose and expression.

% \subsection{Characterizing Rendering Steps of Gaussian-based AR/VR Applications}
\subsection{Workload Summary of Gaussian-Based AR/VR}
\label{pre:pipeline}

From the aforementioned rendering pipelines, we observe that \underline{(1)} diverse Gaussian-based AR/VR applications differ primarily in \textit{Rendering Step \ding{182}}. For instance, they may introduce additional transformations for the geometric parameters of 3D Gaussians or replace spherical harmonics functions with time-conditioned parameterizations; and \underline{(2)} after \textit{Rendering Step \ding{182}}, all rendering pipelines involve the same set of computations in \textit{Rendering Step \ding{183}} and \textit{Rendering Step \ding{184}}.
This observation holds for even more 3D Gaussian-based applications, such as simulating driving scenes~\cite{zhou2024drivinggaussian} and rendering language-embedded semantic images~\cite{qin2023langsplat}.

\section{Profiling and Analysis}

\label{sec:profiling}

\label{sec:profiling:overall}

\begin{figure}[!t]
\centering
\vspace{-0em}
\includegraphics[width=1.0\linewidth]{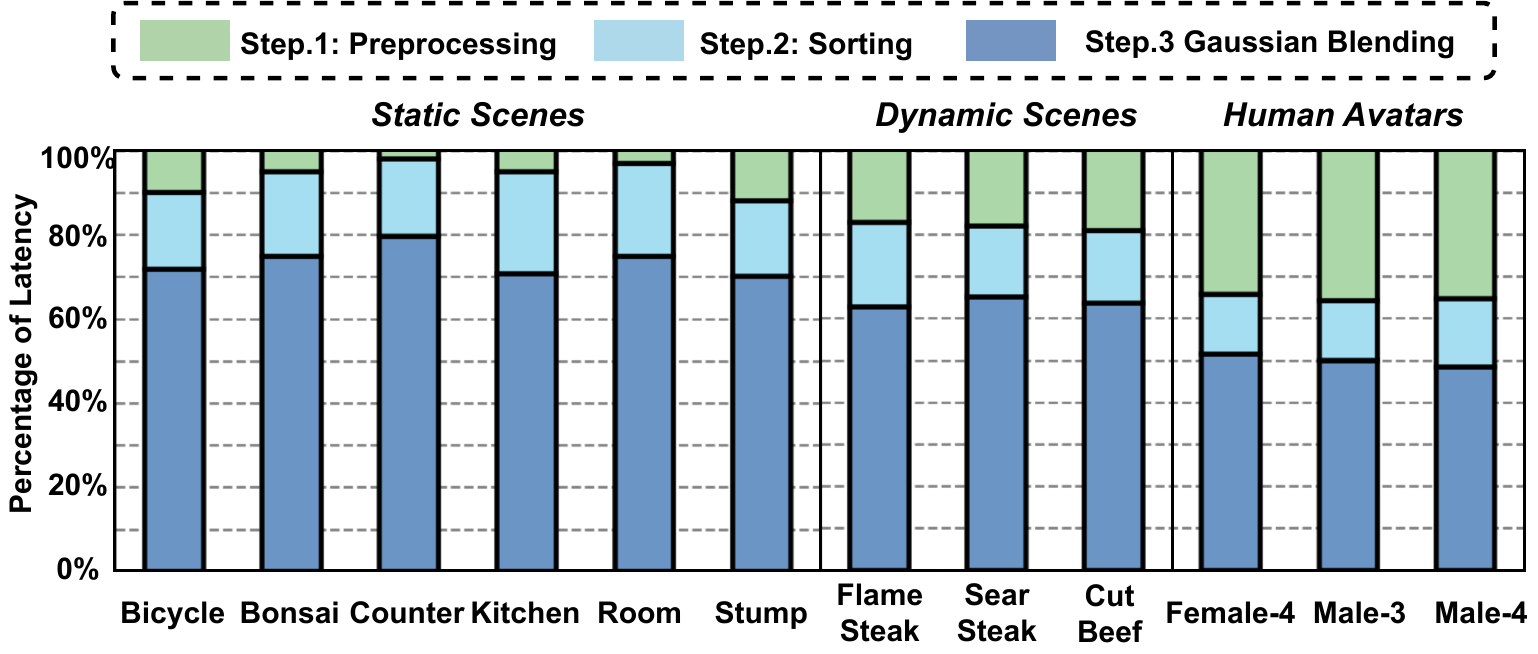}
\vspace{-2em}
\caption{Rendering time breakdown on three real-world datasets~\cite{barron2021mipnerf, Li_2022_CVPR ,alldieck2018video}.} 
\label{fig:profiling_breakdown_2}
\vspace{-1em}
\end{figure}

\begin{table}[b!]
\vspace{-2em}
  \caption{Algorithm and Dataset Setup for Profiling.}
  
  \centering
    \resizebox{1\linewidth}{!}
    {
      \begin{tabular}{c|c|c}
      
      \toprule[1pt]
       \textbf{Scene Type} & \textbf{Scenes} & \textbf{Resolution} \\
      \midrule
      \midrule
      \multirow{2}{*}{Static Scene~\cite{barron2021mipnerf}} &  Bicycle, Bonsai, Counter,  & 779 $\times$ 519 to\\
       &Kitchen, Room, Stump & 1245 $\times$ 825\\
      \midrule
      \multirow{2}{*}{Dynamic Scene~\cite{Li_2022_CVPR}} & flame steak, sear steak, &  \multirow{2}{*}{1352 $\times$ 1014}\\
      & cut beef & \\
      \midrule
       Human Avatar~\cite{alldieck2018video} & female-4, male-3, male-4 & 1080 $\times$ 1080\\
      \bottomrule[1pt]
      \end{tabular}
      }
    \label{tab:dataset}
\end{table}

% \noindent\textbf{Algorithm and Dataset} 
% To understand the performance of Gaussian-based AR/VR applications, we profiled the rendering pipelines of three 3D Gaussian-based reconstruction algorithms, 3D Gaussian Splatting~\cite{kerbl3Dgaussians} for reconstructing static scenes, 4D Gaussian Splatting~\cite{yang2023gs4d} for dynamic scene reconstruction, and Splatting Avatar~\cite{SplattingAvatar} for human avatar animation. Our profiling is conducted on top of real-world datasets~\cite{barron2021mipnerf, Li_2022_CVPR ,alldieck2018video}. The detailed statistics of the datasets are listed in Tab.~\ref{tab:dataset}.

To understand the typical acceleration bottlenecks of Gaussian-based rendering pipelines, we profile popular Gaussian-based reconstruction algorithms on AR/VR platforms. These include 3D Gaussian Splatting~\cite{kerbl3Dgaussians} for reconstructing static scenes, 4D Gaussian Splatting~\cite{yang2023gs4d} for dynamic scene reconstruction, and Splatting Avatar~\cite{SplattingAvatar} for human avatar animation.
Our profiling is conducted using real-world datasets~\cite{barron2021mipnerf, Li_2022_CVPR, alldieck2018video}. The detailed statistics of the datasets are listed in Tab.~\ref{tab:dataset}. We run the algorithms on an edge GPU device, the Jetson Orin NX 16GB~\cite{JetsonOr73:online}, and use \textit{Nisight Systems}~\cite{NsightSy30:online} for a kernel-level rendering time breakdown.

\subsection{Overall Profiling Results}

% We profiled the end-to-end rendering process on the edge GPU, Fig.~\ref{fig:profiling_breakdown_1} provides the overall runtime on 12 real-world scenes of 3 different types: 6 are static scenes, 3 are dynamic scenes, and 3 are human avatars. Fig.~\ref{fig:profiling_breakdown_2} shows the rendering time breakdown into the 3 rendering stages.

% As shown in Fig.~\ref{fig:profiling_breakdown_1} and Fig.~\ref{fig:profiling_breakdown_2}, on the edge GPU, none of the three types of scenes achieves real-time rendering performance ($>$ 60 FPS), this is mainly because of the performance bottleneck in \textbf{Rendering Step.~3}. For example, on the real-world static scenes, this step takes 69.56\% to 78.25\% of the overall rendering time. For dynamic scene reconstruction and animatable avatar, which requires more complex preprocessing for modeling motion and deformation, \textbf{Rendering Step.~1} occupies a larger portion of rendering time. However, \textbf{Rendering Step.~3} is still the major bottleneck, accounting for 61.98\% to 64.85\% of rendering time for dynamic scenes and  47.97\% to 51.19\% of rendering time for human avatar animation. Moreover, \textbf{Rendering Step.~2}, which involves the depth sorting process, also consumes a significant portion of the rendering time across all three types of scenes, ranging from 13.74\% to 24.47\%. 

We summarize the overall runtime and the corresponding rendering time breakdown into the three aforementioned rendering stages in Fig.~\ref{fig:profiling_breakdown_1} and Fig.~\ref{fig:profiling_breakdown_2}, respectively. This profiling encompasses 12 real-world scenes of three different types: 6 static scenes, 3 dynamic scenes, and 3 human avatars.

We observe that \underline{(1)} on the edge GPU, none of the three types of scenes achieves real-time rendering performance ($\geq$ 60 FPS~\cite{framerate}). This is primarily due to the latency bottleneck in \textit{Rendering Step \ding{184}}. For instance, in real-world static scenes, this step accounts for 70\% to 78\% of the overall rendering time; \underline{(2)} In dynamic scenes and animatable avatar rendering, although the percentage of \textit{Rendering Step \ding{182}} increases due to more complex preprocessing steps for modeling motion and deformation, \textit{Rendering Step \ding{184}} remains the major bottleneck, accounting for 62\% to 65\% in dynamic scenes and 48\% to 51\% in human avatar animation; \underline{(3)} \textit{Rendering Step \ding{183}}, which involves the sorting process, also consumes a non-negligible portion of the rendering time across all three types of scenes, ranging from 14\% to 24\%.

\subsection{\zy{Identified Challenges}}
\label{sec:profiling:challenegs}

\begin{figure}[t]
\centering
\vspace{1em}
\includegraphics[width=1.0\linewidth]{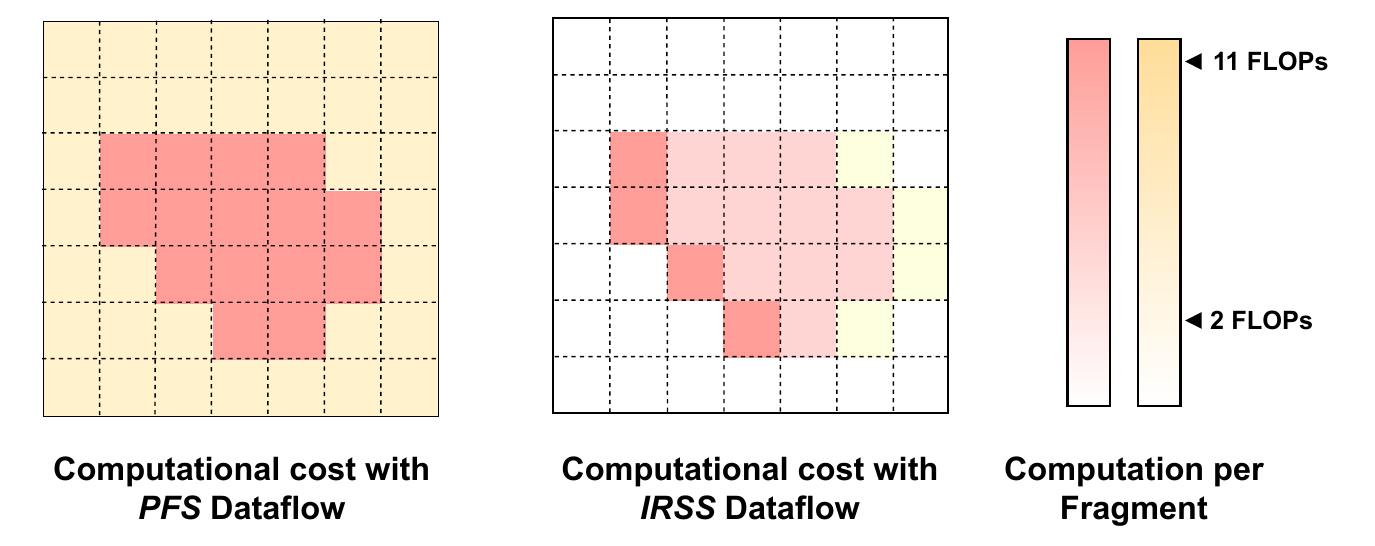}
\vspace{-1.5em}
% \caption{The rendering pipeline of 3D Gaussians.}
\caption{\zy{Comparison of the computational complexity between the original \textit{PFS} dataflow~\cite{kerbl3Dgaussians} and the proposed \textit{IRSS} dataflow. The color depth indicates the per-fragment workload: red represents useful computations on fragments that significantly contribute to the output image, while yellow represents the workload on redundant fragments.}}
\label{fig:compare_dataflow}
\vspace{-1.5em}
\end{figure}

Based on the profiling results, we conducted an in-depth analysis to identify the challenges in accelerating the bottleneck \textit{Rendering Step \ding{184}}.

\noindent\textbf{Challenge 1: Excessive Per-Fragment Computation.} As identified in Sec.~\ref{sec:profiling:overall}, \textit{Rendering Step \ding{184}} is the primary latency bottleneck, consistently consuming more rendering time compared to the other two steps. Unlike the other steps, where computational complexity is determined by the number of Gaussians, \textit{Rendering Step \ding{184}} involves per-fragment computation. Our profiling shows that the average fragment-to-Gaussian ratio is 541:1, 161:1, and 688:1 across the three types of applications\cite{barron2021mipnerf, Li_2022_CVPR, alldieck2018video}, respectively, leading to significantly higher computational complexity in \textit{Rendering Step \ding{184}} as compared to the other steps. Moreover, the per-fragment computation in \textit{Rendering Step \ding{184}} requires multiple matrix-vector multiplications in the exponent of Eq.~\ref{eq:alpha}:
\begin{align}
(P-{\mu}^\star_i)^T{{\Sigma}^\star_i}^{-1}(P-{\mu}^\star_i),
\label{eq:exponent}
\end{align}
amounting to 11 FLOPs per fragment. For real-world static scene rendering~\cite{barron2021mipnerf}, Eq.~\ref{eq:exponent} alone would require 1.1 TFLOPs to achieve 60 FPS, which is 58\% of Jetson Orin NX's peak floating-point throughput~\cite{JetsonOr73:online}.

    \noindent\textbf{Challenge 2: Fragment-Level Redundancy.} Although a 2D Gaussian is sampled, on average, at hundreds of fragments, only 7.6\%, 13.7\%, and 9.9\% of fragments make a non-negligible contribution (i.e., opacity greater than a predefined threshold) to the output colors on the three types of applications~\cite{barron2021mipnerf, Li_2022_CVPR, alldieck2018video}. The high redundancy is associated with the tile-based rendering approach. Specifically, during runtime, each $16 \times 16$ tile is assigned to a Streaming Multiprocessor (SM), and the opacity computation and $\alpha$-blending are conducted in lockstep for all fragments in a tile in parallel. However, a 2D Gaussian function usually significantly contributes to only a portion of the fragments within a $16 \times 16$ tile. Therefore, although this lockstep computation (referred to as \textit{Parallel Fragment Shading (PFS)} dataflow in Sec.~\ref{sec:dataflow}) makes use of the Single Instruction Multiple Thread (SIMT) computational capability of SMs, a large portion of computation is wasted on fragments with negligible contribution.

% \section{Gaussian Blending Unit: Algorithm \textcolor{blue}{(Ready for Check)}}
% \label{sec:algorithm}
\section{The Proposed IRSS Dataflow}
\label{sec:dataflow}

\begin{figure}[!t]
\centering
\vspace{1em}
\includegraphics[width=1.02\linewidth]{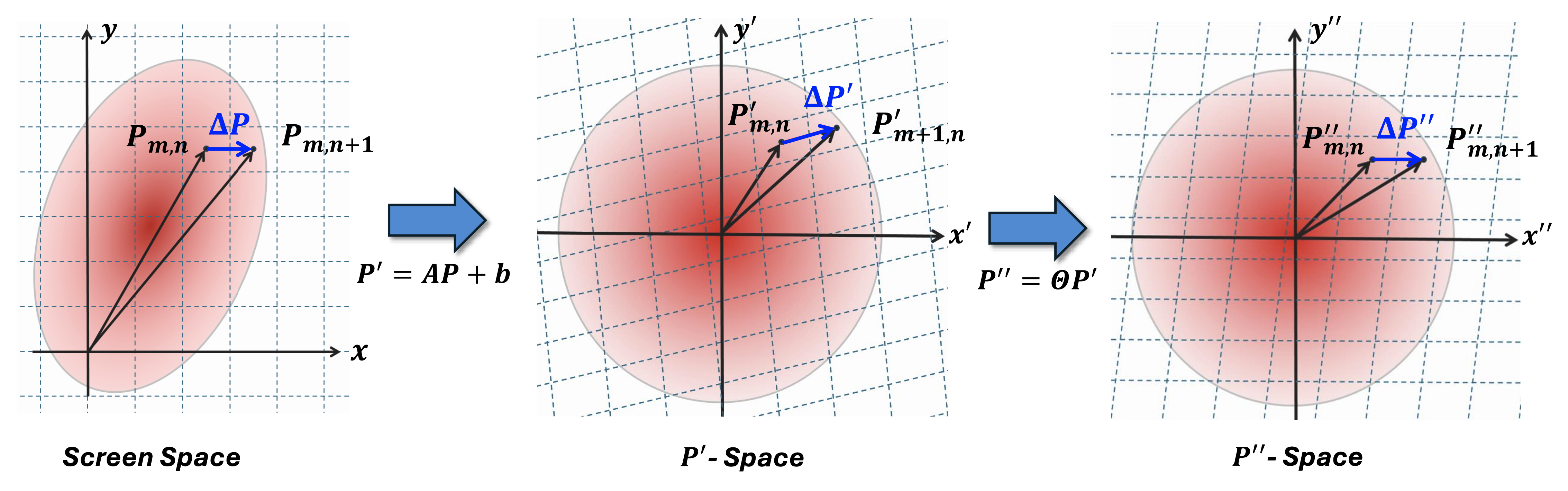}
\vspace{-1.5em}
% \caption{The rendering pipeline of 3D Gaussians.}
\caption{\zy{Illustration of the proposed two transformations that enable compute sharing. In both the $P'$-space and $P''$-space, the squared distance of a fragment from the origin corresponds to \cready{the value of} Eq.~\ref{eq:exponent} in the original screen space. In the $P''$-space, the distance vector between two adjacent fragments in a row, $\Delta P''$, is parallel to the $x''$-axis.}}
\label{fig:transform}
\vspace{-1em}
\end{figure}

\begin{figure*}[!ht]
\centering
\vspace{1em}
\includegraphics[width=1.0\linewidth]{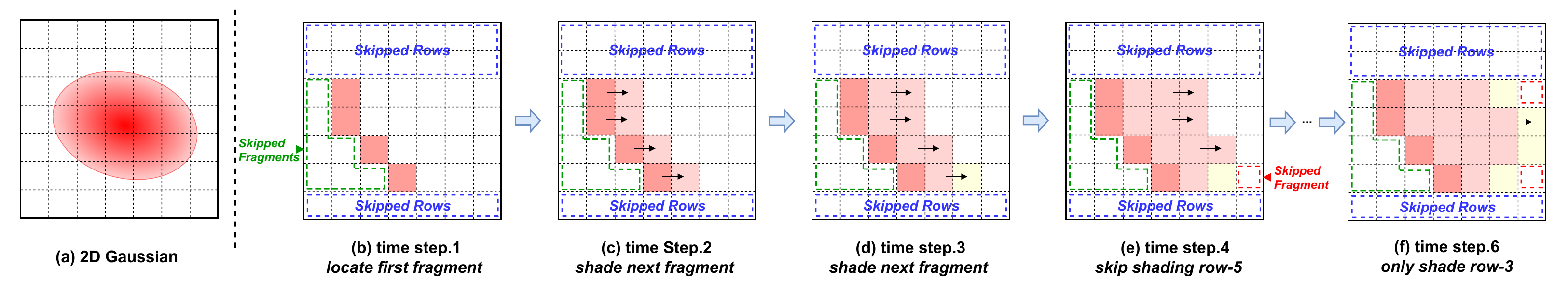}
\vspace{-1.5em}
% \caption{The rendering pipeline of 3D Gaussians.}
\caption{\zy{Illustration of the proposed \textit{IRSS} dataflow: (a) the 2D Gaussian to be rendered; (b) - (f) sequential rendering steps. In (b), during the first time step, fragments in the green zone are skipped as they fall outside the thresholded 2D Gaussian, and the blue box shows skipped rows that are also identified and omitted. In (c) - (f), the black arrow indicates compute sharing between adjacent fragments, resulting in lower computational complexity for newly shaded fragments compared to those shaded in (b). In (d), the last fragment of the 6th row is identified, allowing the computation for this row to be skipped in the next time step (red box in (e)).}}
\label{fig:dataflow}
\vspace{-1em}
\end{figure*}

\subsection{\zy{IRSS Dataflow: Motivation and Overview}}
\label{sec:dataflow:motivation}

% Motivated by the \textbf{Challenge 1} and \textbf{Challenge 2}, we propose an \textit{Intra-Row Sequential Shading (IRSS)} dataflow to improve the efficiency of \textit{Rendering Step \ding{184}}. As illustrated in Fig.~\ref{fig:compare_dataflow}, this \textit{IRSS} dataflow reduces the computational cost compared to the original \textit{PFS} dataflow by sequentially shading (i.e. performing the computation of \textit{Rendering Step \ding{184}} on) the fragments in a row. This computation sharing is non-trivial due to the multi-step geometry transformations required, as detailed in Fig.~\ref{fig:dataflow}:
Motivated by the identified \textbf{Challenge 1} and \textbf{Challenge 2}, we propose an \textit{Intra-Row Sequential Shading (IRSS)} dataflow to improve the efficiency of \textit{Rendering Step \ding{184}}. As illustrated in Fig.~\ref{fig:compare_dataflow}, this \textit{IRSS} dataflow reduces the computational cost compared to the original \textit{PFS} dataflow by sequentially shading (i.e. performing the computation of \textit{Rendering Step \ding{184}} on) the fragments in a row, which enables compute sharing and skip redundant fragments:

\noindent \textbf{Compute Sharing.} By shading the fragments in a row sequentially from left to right, we can share intermediate values among adjacent fragments, reducing the computational complexity of Eq.~\ref{eq:exponent} from 11 FLOPs per fragment to 2 FLOPs per fragment. This computation sharing is non-trivial due to the multi-step geometry transformations needed as detailed in Sec.~\ref{sec:dataflow:compute_sharing} and Fig.~\ref{fig:transform}.

\noindent \textbf{Redundancy Skipping.} The \textit{IRSS} dataflow also facilitates the efficient identification of redundant fragments and rows that contribute negligibly to the output image, allowing us to skip the associated computation. This is accomplished by leveraging the convex shape~\cite{zwicker2001ewa} of a truncated 2D Gaussian. Once the first and last fragments in a row that intersect the Gaussian are identified, all fragments outside this range can be skipped. The detailed implementation is elaborated in Sec.~\ref{sec:dataflow:redundancy_skipping}.

\subsection{IRSS Dataflow: Compute Sharing}
\label{sec:dataflow:compute_sharing}

% Our row-wise dataflow allows computational sharing in rendering stage 3, this is achieved by sharing the intermediate results between adjacent fragments when computing Eq. 5. However, such a computational sharing is non-trivial and involves multiple-step geometry transformation for obtaining maximum computational reduction. Note that the transformations by no means approximation for Eq.5, and the rendering quality is not compromised, as evidenced in Sec. xxx.

% \noindent \textbf{Overview} In response to \textbf{Challenge 1}, the proposed \textit{IRSS} dataflow facilitates the sharing of intermediate values between adjacent fragments shown in Fig.~\ref{fig:dataflow}~(c), thereby reducing computational complexity. However, these sharable intermediate variables are only exposed after appropriate geometric transformations illustrated in Fig.~\ref{fig:transform}. A two-step transformation is required to achieve maximum computation reduction. The two transformations are elaborated below. It is important to note that these transformations are not approximations for Eq.~\ref{eq:exponent}, and the rendering quality remains uncompromised.
\noindent \textbf{Overview.} In response to \textbf{Challenge 1}, the proposed \textit{IRSS} dataflow facilitates the sharing of intermediate values between adjacent fragments shown in Fig.~\ref{fig:dataflow}(c), thereby reducing computational complexity. However, these sharable intermediate variables are only exposed after appropriate geometric transformations illustrated in Fig.~\ref{fig:transform}. A two-step transformation is required to achieve maximum computation reduction. The two transformations are elaborated below. It is important to note that these transformations are not approximations for Eq.~\ref{eq:exponent}, and the rendering quality remains uncompromised.

\noindent \textbf{Transformation $\mathbf{P\rightarrow P'}$.} To expose the sharable intermediate values, the first transformation converts anisotropic 2D Gaussians into isotropic circles, converting Eq.~\ref{eq:exponent} into a distance between a fragment and the origin (i.e., Gaussian center). This transformation is obtained by performing an eigenvalue decomposition (EVD) on the matrix ${{\Sigma}^\star}^{-1}$ in Eq.~\ref{eq:exponent}:
\begin{align}
&(P-{\mu}^\star_i)^T{{\Sigma}^\star_i}^{-1}(P-{\mu}^\star_i) \\
=&(P-{\mu}^\star)^T Q D^{\frac{1}{2}} D^{\frac{1}{2}} Q^T (P-{\mu}^\star),
\end{align}
where $D$ is the diagonal eigenvalue matrix and $Q$ is the eigenvector matrix. The existence of this decomposition is guaranteed by spectral theory\cite{greub2012linear}, as ${\Sigma}^\star$ is a positive-definite symmetric matrix.

Following the EVD, we derive our first coordinate transformation $P \rightarrow P'$ as $P' = D^{\frac{1}{2}} Q^T (P-{\mu}^\star)$, then:
\begin{align}
(P-{\mu}^\star_i)^T{{\Sigma}^\star_i}^{-1}(P-{\mu}^\star_i) ={P'}^TP' = ||P'||^2_2,
\label{eq:sqaure_alternative}
\end{align}
therefore, computing Eq.~\ref{eq:exponent} is mathematically equivalent to measuring the squared distance $||P'||^2_2$ between the fragment center $P'$ and the origin $O'$.

Because the mapping $P \rightarrow P'$ is affine, the distance vector between any adjacent fragment in a row remains a constant. We denote the distance as $\Delta P' = (\Delta x', \Delta y')^T$. If we have mapped one fragment $P_{M,N}$ at row M, column N to the transformed space $P_{M,N} \rightarrow P'_{M,N}$, it is easy to derive the mapped coordinate of its adjacent fragment on the right:
\begin{align}
P'_{M,N+1} = P'_{M,N} + \Delta P'.
\label{eq:increment_P'}
\end{align}
Thus, there is no need to perform the transformation for every fragment, we can sequentially derive the mapped coordinates of a row of fragments by iteratively adopting Eq.~\ref{eq:increment_P'}.

With this per-row sequential computation, the proposed \textit{IRSS} dataflow reduces the computational cost of Eq.~\ref{eq:exponent} from 11 FLOPs per fragment to \cready{3 FLOPs} per fragment (Eq.~\ref{eq:sqaure_alternative}) for all fragments in a row except the first fragment, which still requires 11 FLOPs. However, each iteration of Eq.~\ref{eq:increment_P'} increments the coordinate in both the $x'$- and $y'$-axis, as $\Delta x'$ and $\Delta y'$ are usually both non-zero, so that for each fragment when computing Eq.~\ref{eq:sqaure_alternative}, both ${x'}^2$ and ${y'}^2$ must be recomputed. We can further reduce the computation cost by limiting $\Delta y'$ to zero using the following transformation.

\noindent \textbf{Transformation $\mathbf{P'\rightarrow P''}$.} To address the computation for computing both the ${x'}^2$ and ${y'}^2$, we introduce an additional transformation (rotation) $P''=\Theta P'$, where $\Theta$ is a rotation matrix. Any rotation does not change the vector length, so we have:
\begin{align}
||P''||^2_2 = ||P'||^2_2,
\end{align}
In other words, the squared distance between a fragment center and the origin in the $P''$-space is also mathematically equivalent to Eq.~\ref{eq:exponent}.

As shown in Fig.~\ref{fig:transform}(c), we choose a $\Theta$ that aligns the distance vector between adjacent fragments in a row parallel to the $x''$-axis:
\begin{align}
\Delta P'' = \Theta \Delta P' = (\Delta x'', 0)^T.
\end{align}

As a result, when computing the squared distance $||P''||^2_2={x''}^2+{y''}^2$, we only need to recompute ${x''}^2$ while ${y''}^2$ stays constant for a row of fragments. Therefore, the computational cost for each fragment is further reduced to 2 FLOPs, except for the first fragment in a row.

\subsection{IRSS Dataflow: Redundancy Skipping}
\label{sec:dataflow:redundancy_skipping}

% \noindent \textbf{Overview} Although techniques in Sec.~\ref{sec:dataflow:compute_sharing} reduce the per-fragment computational cost, the large number of unnecessary fragments, as identified in \textbf{Challenge 2}, can still hinder real-time rendering. To efficiently identify these unnecessary fragments and skip the associated computations, we propose a row-wise redundancy skipping mechanism.
\noindent \textbf{Overview.} Although techniques in Sec.~\ref{sec:dataflow:compute_sharing} reduce the per-fragment computational cost, the large number of unnecessary fragments, as identified in \textbf{Challenge 2}, can still hinder real-time rendering. To efficiently identify these unnecessary fragments and skip the associated computations, we propose a row-wise redundancy skipping mechanism.

This mechanism involves locating the first and last fragment in a row that significantly contributes to the output image and skipping all other fragments. This approach ensures that all significant fragments are retained while all others are skipped because the 2D Gaussian function is convex. Fragments with an opacity higher than the predefined threshold will fall between the intersections of the thresholded (truncated) Gaussian and its row. Next, we illustrate how the first and last fragments can be located in a row-wise manner with the proposed \textit{IRSS} dataflow.

\vspace{0.5em}

\noindent \textbf{Locating the First Fragment.} We adopt a 3-step algorithm to find the first fragment that falls into a truncated 2D Gaussian. This algorithm is facilitated by the two-step transformation in Sec.~\ref{sec:dataflow:compute_sharing}, in the $P''$-space, intersection with a truncated 2D Gaussian is determined by whether the fragment coordinate falls inside a 2D circle (i.e., ${x''}^2 + {y''^2} < Th$), where $Th$ is derived by the predefined truncation threshold.

\noindent \underline{Step-1} Obtain the $x''$ and $y''$ for the leftmost fragment in a row. If ${y''}^2 > Th$, then this row has no intersection with this Gaussian and can be skipped entirely because $y''$ is constant for a row of fragments, which is the blue box in Fig.~\ref{fig:dataflow}(b).

\noindent \underline{Step-2} If ${y''}^2 < Th$ and ${x''}^2 + {y''^2} < Th$, then the leftmost fragment is the first fragment that falls into the truncated 2D Gaussian.

\noindent \underline{Step-3} If ${x''}^2 + {y''^2} > Th$, we then check whether $x''$ and $\Delta x''$ have the same sign. If so, no fragments in the current tile intersect with this 2D Gaussian. Otherwise, we perform a binary search to locate the first fragment in this row. All the fragments on the left side of the first fragment are skipped (the green box in Fig.~\ref{fig:dataflow}(b)).

\vspace{0.5em}

\noindent \textbf{Locating the Last Fragment.} Locating the last fragment in a row is straightforward with the proposed \textit{IRSS} dataflow. As we sequentially move right, the first fragment whose ${x''}^2 + {y''^2} > Th$ is the first fragment that falls outside the truncated Gaussian, and all fragments on the right side can be skipped (the red boxes in Fig.~\ref{fig:dataflow}(e) and (f))

\begin{figure}[!t]
\centering
\vspace{1em}
\includegraphics[width=1.0\linewidth]{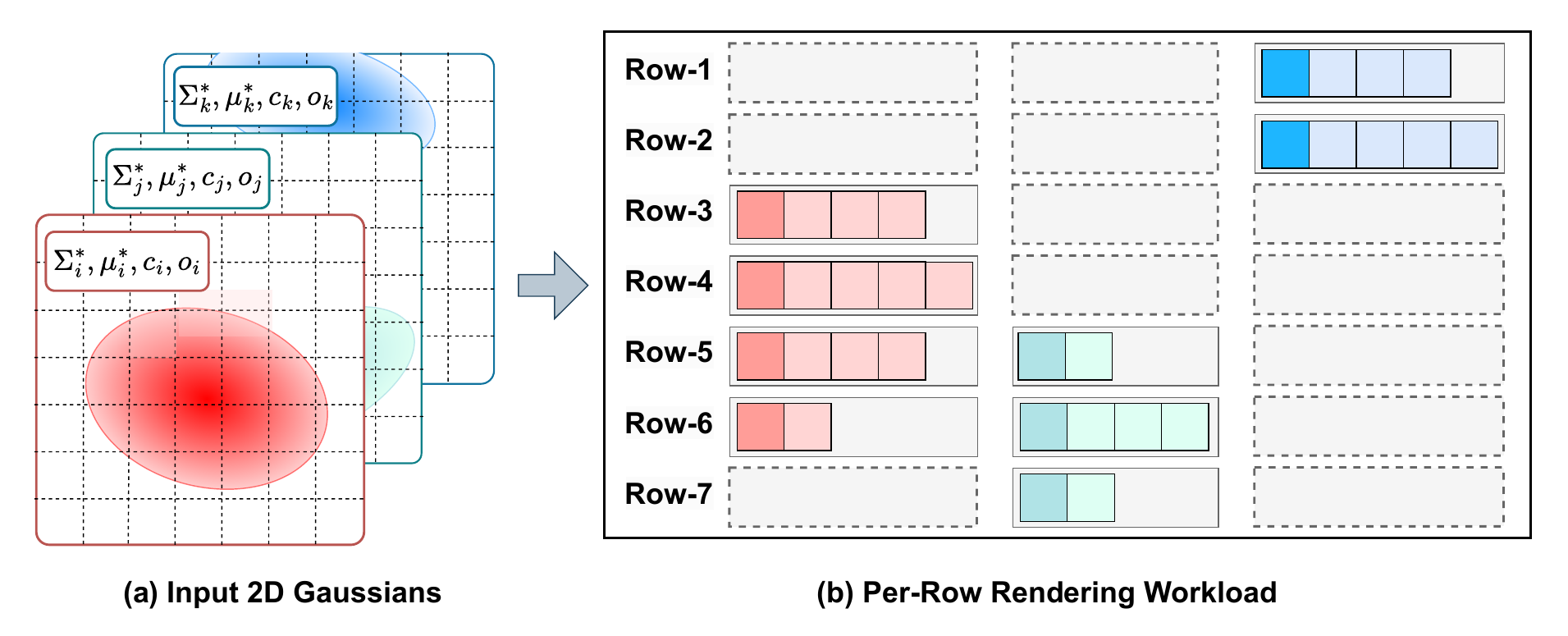}
\vspace{-1.5em}
\caption{Illustration of the input features and per-row workload for \textit{Rendering Step \ding{184}} with the proposed \textit{IRSS} dataflow. One colored block on the right corresponds to one fragment to be shaded.}
\label{fig:hardware_opportunities}
\vspace{-1em}
\end{figure}

\subsection{\zy{Direct Deployment on GPU}}
\label{sec:dataflow:deployment}

We implemented the proposed \textit{IRSS} dataflow as a customized CUDA kernel and benchmarked its performance on the real-world static scene dataset~\cite{barron2021mipnerf}. The experimental results show that the \textit{IRSS} dataflow achieves a significant 59\% reduction in latency during \textit{Rendering Step \ding{184}}, increasing the rendering speed from 13 FPS to 22 FPS. However, it still falls short of meeting the real-time rendering performance requirements for AR/VR applications~\cite{framerate}. In the next section, we identify the limitations of GPU performance and propose a dedicated hardware module, dubbed GBU, to overcome this performance bottleneck.

\vspace{1em}
\section{The Proposed Gaussian Blending Unit}
\label{sec:arch}

\begin{figure}[!t]
\centering
% \vspace{-2.5em}
\vspace{1em}
\includegraphics[width=1.0\linewidth]{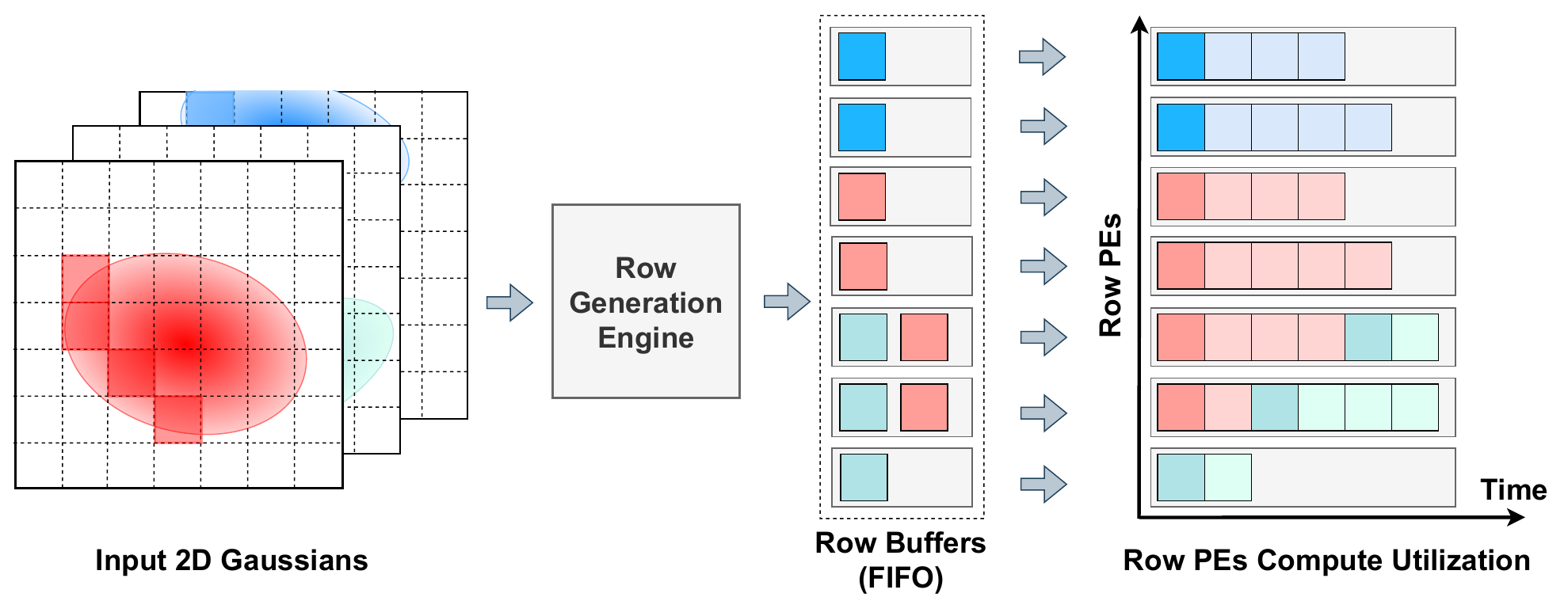}
\vspace{-1em}
\caption{Comparing GPU and GBU hardware utilization when executing the proposed \textit{IRSS} dataflow.}
\label{fig:hardware_amortize}
\vspace{-1em}
\end{figure}

\subsection{\zy{Motivating Profiling}}
\label{sec:arch:motivation}

As discussed in Sec.~\ref{sec:dataflow:redundancy_skipping}, directly deploying the proposed \textit{IRSS} dataflow on a GPU still falls short of achieving real-time rendering performance. Our profiling of the GPU-based implementation on the Jetson Orin NX~\cite{JetsonOr73:online} reveals two limitations for further speedup:

\noindent \textbf{Limitation 1: Low Compute Utilization.} Further rendering speedup is primarily hindered by limited compute utilization on GPUs. As illustrated in Fig.~\ref{fig:hardware_opportunities}(b), after skipping redundant fragments, the compute workloads become heavily imbalanced among different rows. When these rows are mapped to synchronized SIMT threads in a GPU warp, this imbalance results in only 18.9\% utilization of GPU threads/lanes on the real-world static scene dataset~\cite{barron2021mipnerf}.

\noindent \textbf{Limitation 2: High Memory Footprint.} The memory footprint for reading Gaussian features (Fig.~\ref{fig:hardware_opportunities}(a)) in the \textit{Rendering Step \ding{184}} can negatively impact the throughput of the first two rendering steps when the three steps are pipelined. Our profiling of real-world static scenes~\cite{barron2021mipnerf} shows that \textit{Rendering Step \ding{184}} alone requires 62.1\% of DRAM bandwidth to achieve real-time (60 FPS) rendering performance. The experiments in Sec.~\ref{sec:exp:ablation} indicate that this limitation could lead to a 13.5\% slowdown in end-to-end rendering.

\subsection{Hardware System Overview}

\begin{figure*}[!t]
\centering
\vspace{1em}
\includegraphics[width=1.0\linewidth]{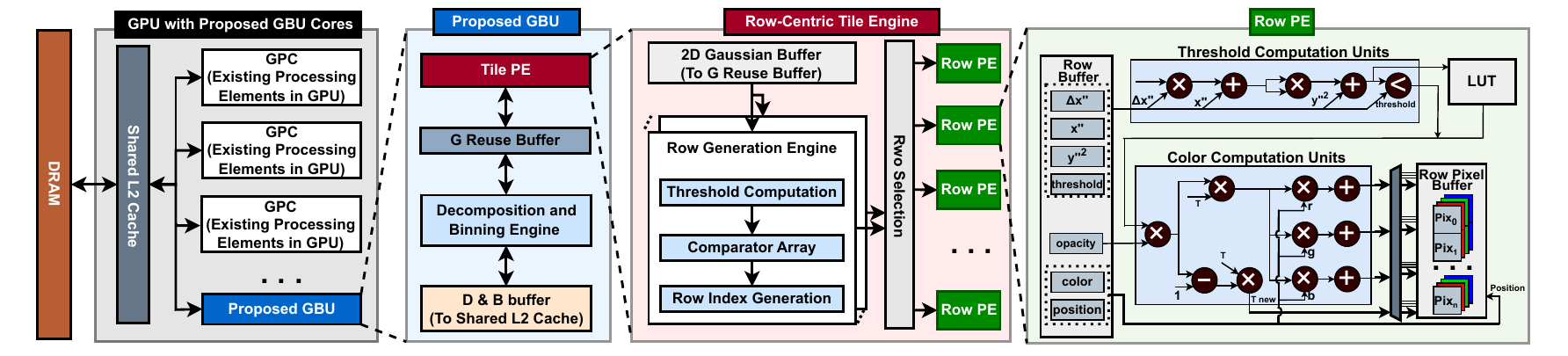}
\vspace{-1em}
\caption{Overview of the proposed hardware acceleration system integrating an edge GPU with the GBU. (a) shows the overall system; (b), (c), and (d) provide zoomed-in views of the GBU, the Tile Engine, and the Row PE, respectively. }
\label{fig:architecture}
\vspace{-0em}
\end{figure*}

In response to the identified limitations, we propose a dedicated acceleration system for Gaussian-based rendering. As illustrated in Fig.~\ref{fig:architecture}(a), this system comprises two key components: \underline{(1)} a Gaussian Blending Unit (GBU) featuring a Row-Centric Tile Engine and a Gaussian Reuse Cache to address \textbf{Limitation 1} and \textbf{Limitation 2}, respectively; and \underline{(2)} seamless integration with GPU architectures. The GBU is deployed outside the GPU's Graphics Processing Clusters (GPC) to enable independent execution and pipelining between the GPU and GBU. With careful workload assignment and pipelining, this integration allows our hardware acceleration system to support a wide range of AR/VR applications.

In the following sections, we introduce the Row-Centric Tile Engine (Sec.~\ref{sec:arch:tile_engine}) and the Gaussian Reuse Cache (Sec.~\ref{sec:arch:cache}). We then elaborate on integrating the proposed GBU into commercial GPU devices ((Sec.~\ref{sec:arch:integration}) and discuss the programming model (Sec.~\ref{sec:arch:api}) for the GBU.

\subsection{GBU: Row-Centric Tile Engine}
\label{sec:arch:tile_engine}

Motivated by the \textbf{Limitation 1}, our per-tile rendering engine, the Row-Centric Tile Engine (shown in Fig.~\ref{fig:architecture}(c)), is designed to maximize compute utilization of the proposed \textit{IRSS} dataflow. This engine renders the $16 \times 16$ image tiles one by one. Instead of shading all rows in a tile with synchronized SIMT lanes, which suffer from low compute utilization due to the imbalanced workload between rows, our tile engine assigns each row to a Row PE. Each Row PE balances the workload by aggregating tasks from multiple Gaussians. As shown in Fig.~\ref{fig:hardware_amortize}, this is supported by (1) a Row Generation Engine, which identifies the fragments to be shaded for each row and forwards them to the corresponding Row Buffer; and (2) a set of Row PEs that consistently poll the fragments to be rendered from the Row Buffer.

\noindent \textbf{Row Generation Engine.} The Row Generation Engine determines which rows a Gaussian intersects and locates the first fragment for each row. Once identified, the position of the first fragment, along with the Gaussian’s color, opacity, truncation threshold, and sharable intermediate values (${y''}^2$, $x''$, and $\Delta x''$) are forwarded to the corresponding Row PE’s Row Buffer.

% 2024.08.01
\noindent \textbf{Row PE.} Each Row PE consists of a Row Buffer, a Threshold Computation Unit, a Color Computation Unit, and a Row Pixel Buffer. The Row Buffer receives the input Gaussian features and the first fragment of the Gaussian in the row. The Threshold Computation Unit and Color Computation Unit then update the accumulated pixel color following the proposed \textit{IRSS} dataflow. To maximize output reuse, the accumulated pixel colors are kept stationary in the Row Pixel Buffer.

\begin{figure}[!t]
    \centering
    \includegraphics[width=1.00\linewidth]{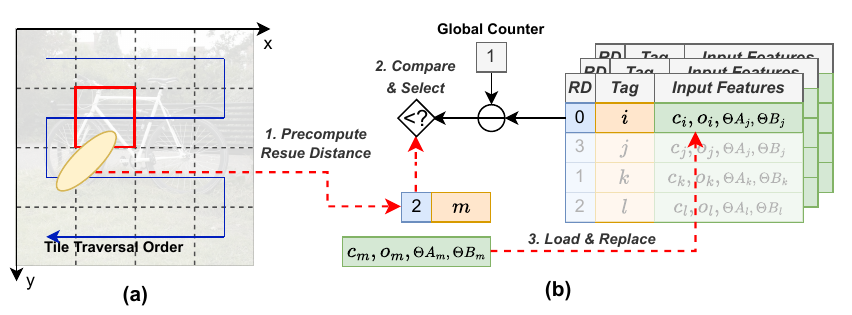}
    \vspace{-1.6em}
    \caption{(a) Pre-computing the reuse distance of input Gaussian features, performed by the Decomposition and Binning Engine, which adapts the algorithm in Sec.~\ref{sec:dataflow:redundancy_skipping} for Gaussian-tile intersection test. (b) Demonstration of a cache replacement, detailed in Sec.~\ref{sec:arch:cache}.}
    \label{fig:morton}
    \vspace{-1.5em}
\end{figure}

\subsection{GBU: Gaussian Reuse Cache}
\label{sec:arch:cache}

To minimize off-chip memory accesses for input Gaussian features (\textbf{Limitation 2}), we propose a Gaussian reuse cache that enhances feature reuse. We leverage a key insight that the access sequence for input Gaussian features can be precomputed. This allows us to implement an optimized cache replacement policy to maximize the opportunity for feature reuse. As shown in Fig.~\ref{fig:morton}, our cache replacement policy has the following four steps:

\noindent \underline{\textit{Step 1: Precompute Reuse Distance}} The reuse distance of an input Gaussian feature is defined as the number of tiles processed before the feature is accessed again by the tile engine. This distance can be precomputed by testing which tiles a Gaussian intersects, as shown in Fig.~\ref{fig:morton}(a). A dedicated Decomposition and Binning engine performs this Gaussian-tile intersection test, generating a list of intersected Gaussians with their corresponding reuse distances for each tile.

\noindent \underline{\textit{Step 2: Compare \& Select}} On a cache miss, the tile engine selects the Gaussian feature with the longest reuse distance by comparing the RD (reuse distance) fields of the Gaussian features. The reuse distances of all Gaussian features at the current tile are computed by subtracting a global counter that tracks the number of processed tiles.

\noindent \underline{\textit{Step 3: Load \& Replace}} On a cache miss, the required features are loaded from off-chip memory to replace the selected cache line (Gaussian feature). The global counter increments the RD field before cache installation.

\noindent \underline{\textit{Step 4: Update Reuse Distance}} On a cache hit, the RD field of the corresponding Gaussian feature is updated with the next precomputed reuse distance of the Gaussian plus the global counter.

\subsection{Integration with GPUs for End-to-end Rendering}
\label{sec:arch:integration}

As illustrated in Fig.~\ref{fig:architecture}(a), the proposed GBU is integrated with GPU for efficient end-to-end rendering. This integration follows two design principles: (1) \underline{\textit{Versatility}} to ensure compatibility with a variety of Gaussian-based AR/VR applications and (2) \underline{\textit{Efficiency}} to achieve real-time rendering performance. These principles necessitate careful workload assignment between the GPU and GBU and a two-level pipelined execution.

\noindent \textbf{Workload Assignment.} As analyzed in Sec.~\ref{sec:pre}, \textit{Rendering Step \ding{184}} uses a common algorithm across different Gaussian-based AR/VR applications and consistently acts as the latency bottleneck, while the algorithm in \textit{Rendering Step \ding{182}} varies across applications. To achieve the desired rendering speedup while maintaining compatibility with existing and future Gaussian-based rendering pipelines, we opt to accelerate only \textit{Rendering Step \ding{184}} on the GBU, while offloading the other steps to the highly programmable GPU hardware.

\begin{figure}[!t]
    \centering
    \vspace{-1em}
    \includegraphics[width=0.95\linewidth]{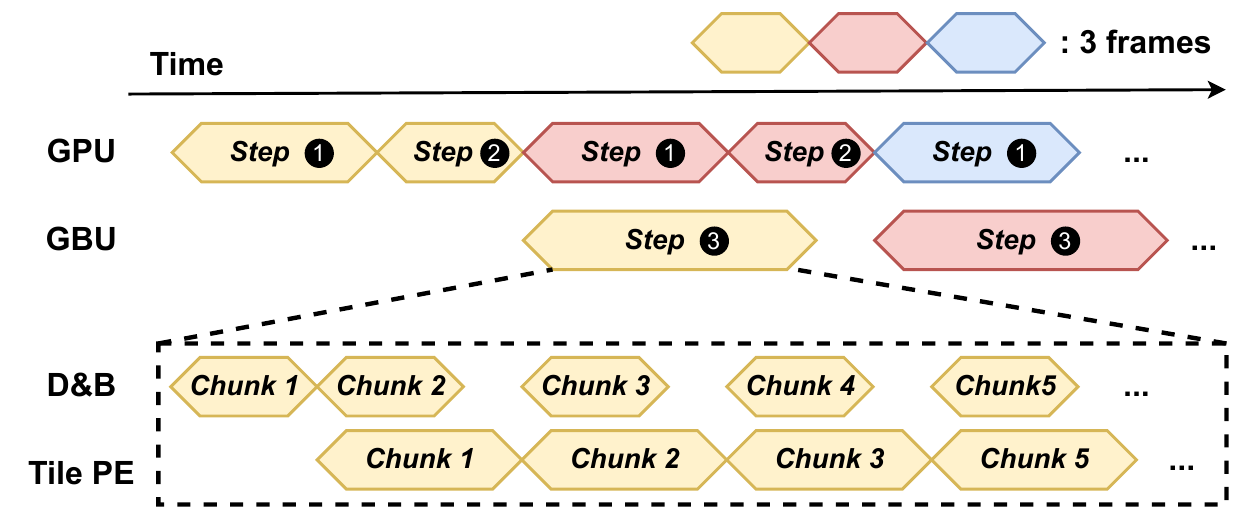}
    \caption{Illustration of the proposed two-level pipeline: top: pipeline between GBU and GPU; bottom: pipeline between Decomposition \& Binning Engine (D\&B) and Tile PE.}
    \label{fig:integration_pipeline}
    \vspace{-1em}
\end{figure}

\noindent \textbf{Two-level Pipeline.} To enhance the overall rendering system's efficiency, we implement a two-level pipeline (illustrated in Fig.~\ref{fig:integration_pipeline}) to increase the utilization of GPU and GBU compute units. The first-level pipeline between the GBU and GPU overlaps \textit{Rendering Step \ding{184}} with the other two rendering steps of the next frame, supported by a pre-allocated double buffer in DRAM. The second-level pipeline parallelizes the execution of the D\&B Engine and the Tile PE by dividing the 2D Gaussians to be rendered into chunks following the depth order. Once one chunk of Gaussians has been assigned (binned) to tiles, the Tile PE can start the rendering process for that chunk.

\subsection{Programming Model}
\label{sec:arch:api}

% \begin{figure}[!h]
%     \centering
%     \includegraphics[width=1.0\linewidth]{Figures/API.png}
%     \vspace{-1.5em}
%     \caption{C programming interface of the proposed Gaussian Blending Unit.}
%     \label{fig:api}
%     % \vspace{-2em}
% \end{figure}

GBU's programming model (Listing.~\ref{listing:api}) is designed to offer full flexibility and control, making it easy to utilize GBU for accelerating various AR/VR applications. GBU provides two function calls:

\begin{lstlisting}[caption=C++ programming interface of GBU.,language=C++,frame=single,label=listing:api]
void GBU_render_image(
    int H, int W, // image height and width
    const void* input_feature, // Gaussian features
    const unsigned* sorted_index, // depth order
    void* frame_buffer, // buffer for output image
    int ch=3, // number of color channels
);

int GBU_check_status(
    bool blocking //whether to wait till complete
);
\end{lstlisting}
\vspace{1em}

\noindent \textbf{GBU\_render\_image.} This function triggers GBU to render a single image. It takes as input a pointer to the output of the first two rendering steps and writes the rendering output to a preallocated frame buffer. The color channel is configurable and is set to 3 by default.

\noindent \textbf{GBU\_check\_status.} This function returns the execution status of GBU: 0 (idle) or 1 (in execution). It also includes an optional blocking flag to block a CPU thread until GBU becomes idle. GBU does not automatically synchronize with any CUDA streams and depends on this function to implement the aforementioned GBU-GPU pipeline.

% \noindent \textbf{GBU\_render\_image} This function performs \textit{Rendering Step \ding{184}} for a frame. It takes the pointer to the output of the first two rendering steps and writes the rendering output to a pre-allocated frame buffer. The execution of GBU does not lie on any CUDA Streams, so this function will immediately start GBU execution.

% \noindent \textbf{GBU\_check\_status} To implement the aforementioned pipeline execution between GBU and GPU, we provide a \textit{GBU\_check\_status} that checks the completion status of GBU, it also provides an optional blocking flag to block CPU execution until GBU writes its output to the frame buffer.

\vspace{1em}
\section{Experimental Results}
\label{sec:exp}

\subsection{Experiment Setup}
\label{sec:exp:setup}

\begin{table}[t!]
      \vspace{-1em}
  \caption{Specification of GBU and Jetson Orin NX.}
  
  \centering
    \resizebox{1\linewidth}{!}
    {
      \begin{tabular}{l | c c c c c}
      
      \toprule[1pt]
       \multirow{2}{*}{\textbf{Device}} & \multirow{2}{*}{\textbf{SRAM}} & \multirow{2}{*}{\textbf{Area}} & \multirow{2}{*}{\textbf{Frequency}} &\multirow{2}{*}{\textbf{Technology}} & \textbf{Typical}\\
       & & & & & \textbf{Power}\\
      \midrule
      Orin NX~\cite{JetsonOr73:online} & 4 MB& 450 mm$^2$ & 918 MHz & 8 nm & 15 W\\
      \midrule
      GBU & 63 KB & 0.90 mm$^2$ & 1GHz & 28 nm & 0.22 W \\
      \bottomrule[1pt]
      \end{tabular}
      }
      \vspace{-1em}
    \label{tab:hardware}
\end{table}

\begin{table}[t!]
  \caption{Area and power breakdown of GBU hardware modules.}
  
  \centering
    \resizebox{1\linewidth}{!}
    {
      \begin{tabular}{l | c c c c}
      
      \toprule[1pt]
       \textbf{Module} & \textbf{Row PEs} & \textbf{Row Gen.} & \textbf{D\&B Engine} & \textbf{Cache \& Others}\\
      \midrule
      Area (mm$^2$) & 0.36 & 0.14 & 0.10 & 0.30 \\
      \midrule
      Power (W) & 0.11 & 0.04 & 0.03 & 0.04 \\
      \bottomrule[1pt]
      \end{tabular}
      }
      \vspace{-1em}
    \label{tab:hardware_breakdown}
\end{table}

% \noindent\textbf{Datasets and Algorithm.} 
\noindent\textbf{Datasets and Algorithms.} 
Using the same algorithms and datasets as those described in Sec.~\ref{sec:profiling}, we evaluate GBU on 12 real-world scenes from 3 AR/VR applications: 6 scenes for static scene reconstruction~\cite{barron2021mipnerf}, 3 scenes for dynamic scene reconstruction~\cite{Li_2022_CVPR}, and 3 scenes for human avatar animation~\cite{alldieck2018video}. All scenes are real-world captured and the resolution of the scenes ranges from $779 \times 519$ to $1352 \times 1014$, as listed in Tab.~\ref{tab:dataset}. We adopt the following Gaussian-based rendering pipelines for the three types of scenes: the vanilla 3D Gaussian splatting~\cite{kerbl3Dgaussians} for static scene reconstruction, 4D Gaussian splatting~\cite{yang2023gs4d} for dynamic scene reconstruction, and SplattingAvatar~\cite{SplattingAvatar} for animatable human avatars.

\noindent\textbf{Hardware Setup.} We implemented the proposed GBU in Verilog and used Cadence Genus to synthesize the RTL design to gate-level netlist for estimating chip area, timing, and power consumption based on a commercial 28nm CMOS technology. The synthesized frequency is set to 1 GHz. We instantiate one Tile PE on the GPU, which renders a tile (i.e., $16 \times 16$ pixels) at a time. Each Tile PE has 8 Row PEs, and each row PE renders 2 rows inside the tile (i.e. $2 \times 16$ pixels in total). The area and power of GBU and the baselines are presented in Tab.~\ref{tab:hardware} and Tab.~\ref{tab:hardware_breakdown}. We replace one SM on the Jetson Orin NX~\cite{JetsonOr73:online} with GBU and reuse the SM-to-DRAM network to avoid extra area.

\noindent\textbf{Simulation Setup.} For simulating the rendering throughput of a GBU when integrated with an edge GPU, e.g., Jetson Orin NX~\cite{JetsonOr73:online}, we build a cycle-accurate emulator on top of GPGPU-Sim~\cite{khairy2020accel}. For each of the three aforementioned rendering pipelines used in the evaluation, we validate the emulator with measured runtime and power consumption of CUDA kernels in \textit{Rendering Step \ding{182}} and \textit{Rendering Step \ding{183}} on a Jetson Orin NX GPU. The emulated runtime and power consumption are within 10\% error of the real-device measurement.

% \textcolor{blue}{describe how we implement the simulator with GPGPU-Sim, aligning the base configuration with Jetxon NX?}

% \textcolor{blue}{for energy and area modelling, how do we scale everything to 8nm?}

\subsection{Performance on Real-World Scenes}
\label{sec:exp:performance}

\begin{table}[b!]
    \vspace{-1em}
  \caption{Rendering Quality Benchmark.}
  
  \centering
    \resizebox{1\linewidth}{!}
    {
      \begin{tabular}{l | c c |c c |cc}
      
      \toprule[1pt]
      & \multicolumn{2}{c|}{\textbf{Static Scenes~\cite{barron2021mipnerf}}} & \multicolumn{2}{c|}{\textbf{Dynamic Scenes~\cite{li2022neural}}} & \multicolumn{2}{c}{\textbf{Human Avatar~\cite{attal2023hyperreel}}} \\
       & {\textbf{PSNR}$\uparrow$} &{\textbf{LPIPS}$\downarrow$} & {\textbf{PSNR}$\uparrow$} &{\textbf{LPIPS}$\downarrow$} & {\textbf{PSNR}$\uparrow$} &{\textbf{LPIPS}$\downarrow$}\\
      \midrule
      \textbf{3D-GS~\cite{kerbl3Dgaussians}} &  28.90 & 0.196 & 33.80 & 0.976 & 32.19 & 0.022 \\
      \midrule
       \textbf{GBU} & 28.84 & 0.197 & 33.71 & 0.977 & 32.17 & 0.022\\
      \bottomrule[1pt]
      \end{tabular}
}
      \vspace{0em}
    \label{tab:rendering_quality_benchmark}
\end{table}

\begin{table}[b!]
\vspace{-1em}
  \caption{Ablation Study: adding techniques one by one to the acceleration system. }

  \centering
    \resizebox{1\linewidth}{!}
    {
      \begin{tabular}{l | c c c c}
      
      \toprule[1pt]
       & {\textbf{Rendering}} & \textbf{Energy} & \multirow{2}{*}{\textbf{PSNR}$\uparrow$} & \multirow{2}{*}{\textbf{LPIPS}$\downarrow$}\\
       & \textbf{FPS} & \textbf{Efficiency} & \\
      \midrule
      \textbf{Jetson Orin NX~\cite{JetsonOr73:online}} &  12.8 & 1 $\times$ & 28.90 & 0.196 \\
      \midrule
      \hspace{0.5em}+ \footnotesize{IRSS Dataflow} & 22.0  & 1.71 $\times$ & 28.90 & 0.196\\
      \hspace{0.5em}+ \footnotesize{GBU Tile Engine} & 66.1 & 7.22 $\times$ & 28.84 & 0.197\\
      \hspace{0.5em}+ \footnotesize{GBU D\&B Engine} & 80.6 & 9.40 $\times$ & 28.84 & 0.197 \\
      \hspace{0.5em}+ \footnotesize{GBU Reuse Cache} &  91.5 & 10.8 $\times$ & 28.84 & 0.197\\
      \bottomrule[1pt]
      \end{tabular}
}
    \label{tab:hardware_ablation}
\end{table}

\begin{figure}[!t]
\centering
\vspace{0em}
\includegraphics[width=1.0\linewidth]{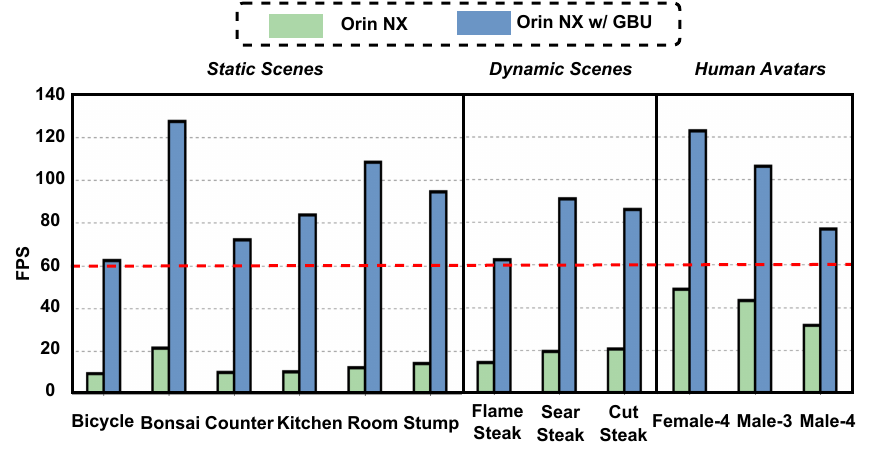}
\vspace{-1.5em}
\caption{Rendering speeds on the 3 different types of tasks on the baseline Jetson Orinx NX GPU~\cite{JetsonOr73:online} and when enhanced with our proposed GBU.}
\label{fig:end-to-end-fps}
\vspace{-1em}
\end{figure}

\noindent\textbf{Rendering Speed.} Fig.~\ref{fig:end-to-end-fps} shows the end-to-end rendering speed of the GBU-enhanced edge GPU compared to the baseline edge GPU. Across all three types of scenes, the proposed Gaussian Blending Unit enables real-time rendering performance (over 60 FPS). On average, the GBU-enhanced edge GPU achieves 92 FPS for static scenes, 80 FPS for dynamic scenes, and 102 FPS for human avatars, while the edge GPU alone only reaches 13 FPS, 18 FPS, and 41 FPS for these respective scenes.

\begin{figure}[t!]
\vspace{1.5em}
\centering
\includegraphics[width=1.0\linewidth]{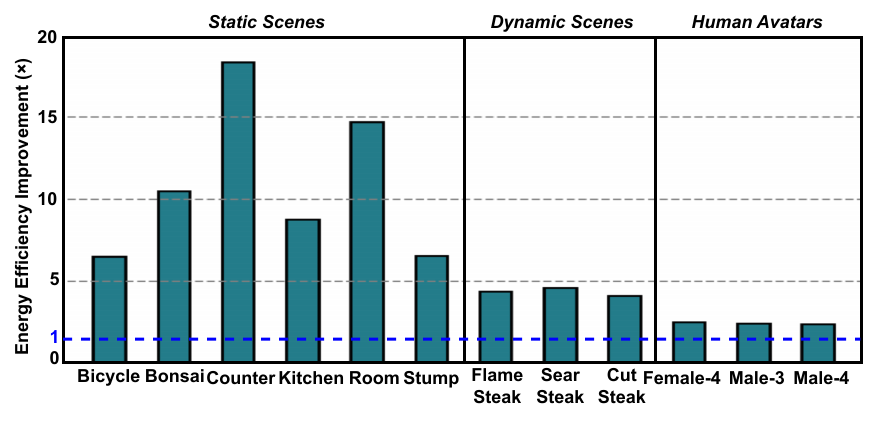}
\vspace{-2em}
\caption{Energy efficiency improvements of our proposed GBU over the baseline Jetson Orinx NX GPU~\cite{JetsonOr73:online}}
\label{fig:energy_efficicy}
\vspace{-0.8em}
\end{figure}

\noindent \textbf{Energy Efficiency.} Fig.~\ref{fig:energy_efficicy} shows the overall energy efficiency improvement. On average, when enhanced with the GBU, energy efficiency improves by 10.8$\times$, 4.4$\times$, 2.5$\times$ on the three types of scenes. This efficiency gain is attributed to our efficient \textit{IRSS} dataflow and optimized hardware implementation, as empirically shown in Sec.~\ref{sec:exp:ablation}. The improvement in energy efficiency for human avatar scenes is lower because these scenes are less bottlenecked by the accelerated \textit{Rendering Step \ding{184}}. As a result, the average energy consumption of rendering 60 images on the three datasets is reduced from 76 J, 52 J, and 23 J to 7 J, 12 J, and 9 J, respectively.

\noindent \textbf{Rendering Quality.} Tab.~\ref{tab:rendering_quality_benchmark} compares the rendering quality between GBU and the original 3D Gaussian Splatting implementation on GPU. We use commonly adopted metrics in the algorithm community to measure rendering quality: Peak Signal-to-Noise Ratio (PSNR, the higher the better) and Learned Perceptual Image Patch Similarity (LPIPS, the lower the better). Across all three types of scenes, GBU hardware only minimally degrades rendering quality ($<$ 0.1 PSNR and $<$ 0.01 LPIPS), which is mainly due to the use of FP-16 precision in the Row-Centric Tile PE. The proposed \textit{IRSS} dataflow itself causes no rendering quality loss when directly deployed on a GPU, as detailed in the ablation study Sec.~\ref{sec:exp:ablation}.

\subsection{Understanding Performance Gains}
\label{sec:exp:ablation}

To understand the rendering speed improvement described in Sec.~\ref{sec:exp:performance}, Tab.~\ref{tab:hardware_ablation} presents the results of an ablation study on the proposed techniques, conducted on real-world static scenes~\cite{barron2021mipnerf}. We observed the following: (1) the proposed \textit{IRSS} dataflow, when directly implemented on a GPU as a customized CUDA kernel, results in a 1.71$\times$ rendering speed boost without compromising rendering quality; (2) integration with the proposed Tile Engine achieves an average of 66.1 FPS, owing to the highly optimized implementation of the Row-Centric Tile Engine. The slight decrease in rendering quality is attributed to the use of 16-bit floating-point precision; (3) the D\&B Engine further increases rendering speed by 1.21$\times$ rendering speed increase by offloading the transformation matrix computation and Gaussian-tile intersection tests from the GPU; (4) the adoption of the Gaussian Reuse Cache, in addition to the Tile Engine and D\&B Engine, further enhances rendering speed by 1.14$\times$, by reducing 44.9\% off-chip memory accesses of \textit{Rendering Step \ding{184}}.

\begin{figure}[!t]
\centering
\vspace{0.5em}
\includegraphics[width=1.0\linewidth]{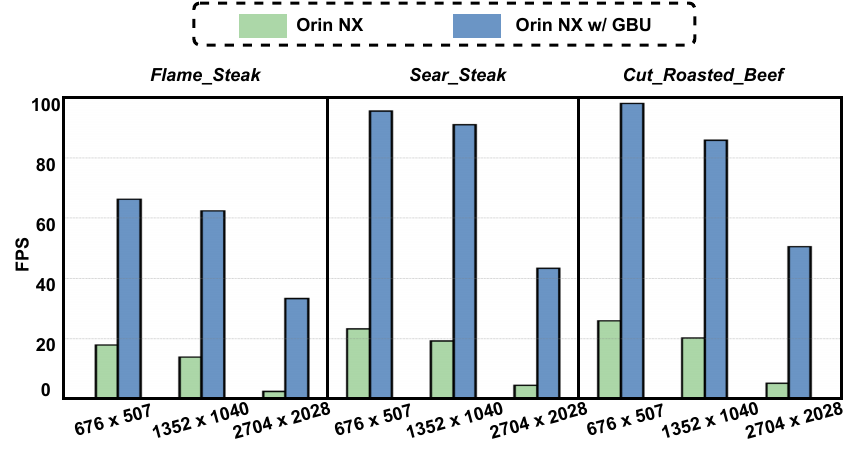}
\vspace{-2em}
\caption{Benchmarking rendering speed of the baseline edge GPU~\cite{JetsonOr73:online} and GBU enhanced edge GPU under different resolutions.}
\vspace{0em}
\label{fig:highers_performance}
\end{figure}

\begin{figure}[!t]
\centering
\vspace{-0em}
\includegraphics[width=1.0\linewidth]{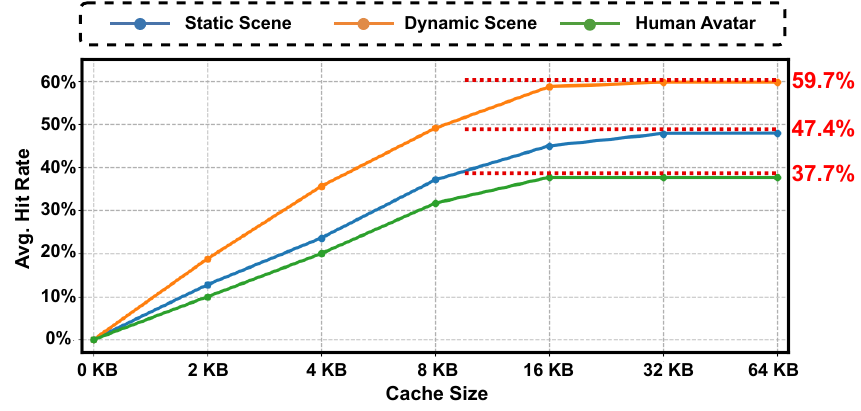}
\vspace{-2em}
\caption{Average hit rates of the Gaussian Reuse Cache across three datasets for varying cache sizes. The dotted red line represents the cache hit rate at a 64 KB cache size.}
\vspace{-1em}
\label{fig:cache_analysis}
\end{figure}

\subsection{Performance Scaling under High Rendering Resolution}

In this section, we analyze the performance of GBU under varying rendering resolutions (from 676 $\times$ 507 to 2704 $\times$ 2028) on the three dynamic scenes~\cite{Li_2022_CVPR}. The rendering speeds are shown in Fig.~\ref{fig:highers_performance}. In particular, GBU achieves a higher acceleration ratio at higher rendering resolutions, e.g. 9.5$\times$ to 13.2$\times$ speedup at 2704 $\times$ 2028 resolution, as compared to 3.7$\times$ to 4.1$\times$ speedup at a resolution of 676 $\times$ 507. This is because the number of fragments grows with the increase in rendering resolution, therefore rendering at a higher resolution exacerbates the bottleneck per-fragment computation in \textit{Rendering Step \ding{184}}. As a result, the dedicated acceleration for \textit{Rendering Step \ding{184}} plays a more vital role in enabling real-time rendering with 3D Gaussians on a higher resolution, making the proposed GBU even more desirable for future-generation AR/VR devices at higher screen resolutions.

\begin{table}[b!]
\vspace{-1em}
  \caption{Specification of GBU-Standlone and the GSCore.}
  
  \centering
    \resizebox{1\linewidth}{!}
    {
      \begin{tabular}{l | c c c | c c}
      
      \toprule[1pt]
       \multirow{2}{*}{\textbf{Device}} & \multirow{2}{*}{\textbf{SRAM}} & \multirow{2}{*}{\textbf{Area}} & \textbf{Typical} & \multicolumn{2}{c}{\textbf{Step \ding{184} PE}}\\
       & & &\textbf{Power} & \textbf{Area} & \textbf{Power} \\
      \midrule
      GS-Core~\cite{lee2024gscore} & 272 KB & 3.95 mm$^2$ & 0.87 W & 1.81 mm$^2$ & 0.25 W \\
      \midrule
      GBU-Standlone & 63 KB & 1.78 mm$^2$ & 0.78 W & 0.50 mm$^2$ & 0.15 W\\
      \bottomrule[1pt]
      \end{tabular}
      }
      \vspace{-0em}
    \label{tab:hardware_standalone}
\end{table}

\subsection{Ablation Study on Cache Sizes}

To understand the impact of cache sizes on cache hit rates, Fig.~\ref{fig:cache_analysis} presents the cache hit rates across varying Gaussian reuse cache sizes (ranging from 0 KB to 64 KB). The results are averaged across the static scene dataset~\cite{barron2021mipnerf}, the dynamic scene dataset~\cite{Li_2022_CVPR}, and the human avatar dataset~\cite{alldieck2018video}, respectively. As shown in Fig.~\ref{fig:cache_analysis}, doubling the cache size results in a linear increase in hit rate when the cache size is below 8 KB. However, the hit rate saturates around 32 KB on all three datasets, with minimal further gains on cache hit rates (i.e., less than $0.1\%$ improvement) beyond this point. Based on this analysis, we configure the Gaussian reuse cache of GBU to be 32 KB.
% Common Question 1: Analysis on cache hit rates.

\subsection{Discussions}

\begin{table}[t!]
% \vspace{-1.5em}

  \caption{{Benchmark with the Previous NeRF Accelerators on the NeRF-Synthetic Dataset~\cite{mildenhall2021nerf}.}}
  
  \centering
    \resizebox{1\linewidth}{!}
    {
      % \begin{tabular}{l | c c c | c c}
      
      % \toprule[1pt]
      %  \multirow{2}{*}{\textbf{Device}} & \multirow{2}{*}{\textbf{Algorithm}} & \multirow{2}{*}{\textbf{Area}} & \textbf{Typical} & \textbf{Typical} & \textbf{Average} \\
      %  & & &\textbf{Power} & \textbf{FPS} & \textbf{PSNR} \\
      % \midrule
      % ICARUS~\cite{rao2022icarus} & NeRF~\cite{mildenhall2021nerf} & N/A & 0.3 W & 0.03 & 30.21 \\      
      % \midrule
      % RT-NeRF-Edge~\cite{li2022rt} & TensoRF~\cite{Chen2022ECCV} & 18.85 mm$^2$ & 8 W & 45 & 31.79 \\
      % \midrule
      % Instant-3D~\cite{li2023instant} & Instant-NGP~\cite{mueller2022instant} & 6.8 mm$^2$ & 1.9 W & $>$ 30 & 33.18* \\
      % \midrule
      % GBU-Standlone & 3D-GS~\cite{kerbl3Dgaussians} & 1.78 mm$^2$ & 0.78 W & 172 & 33.26\\
      % \bottomrule[1pt]
      % \end{tabular}
      % }
    \begin{tabular}{c|cccc}
        \toprule[1pt]
        \textbf{Device} & \textbf{ICARUS~\cite{rao2022icarus}} & \textbf{RT-NeRF~\cite{li2022rt}} & \textbf{Instant-3D~\cite{li2023instant}} & \textbf{GBU-Standalone} \\ \midrule
        \textbf{Algorithm} & NeRF~\cite{mildenhall2021nerf} & TensoRF~\cite{Chen2022ECCV} & Instant-NGP~\cite{mueller2022instant} & 3D-GS~\cite{kerbl3Dgaussians} \\ \midrule
        \textbf{PSNR} & 30.21 & 31.79 & 33.18 & 33.26 \\ \midrule
        \textbf{Technology} & 40nm & 28nm & 28nm & 28nm \\ \midrule
        \textbf{Frequency} & 0.3 GHz & 1.0 GHz & 0.8 GHz & 1.0 GHz \\ \midrule
        \textbf{Area} & N/A & 18.85 mm\(^2\) & 6.8 mm\(^2\) & 1.78 mm\(^2\) \\ \midrule
        \textbf{Power} & 0.3 W & 8 W & 1.9 W & 0.78 W \\ \midrule
        \textbf{FPS} & 0.03 & 45 & $>$ 30 & 172 \\ \midrule
        \bottomrule[1pt]
    \end{tabular}
    }
      \vspace{-1em}
    \label{tab:benchmark_nerf_accelerator}
\end{table}

% \noindent\revi{\textbf{Comparison with NeRF Accelerators} The literature of existing NeRF rendering accelerators~\cite{} mainly designed as standalone accelerators that specialize at rendering static scenes, while the proposed GBU, by leveraging GPUs to handle the distinct \textit{Rendering Step \ding{182}} in different applications, can support a large variety of AR/VR applications~\cite{}. For a fair comparison, we extend GBU as a standalone static scene rendering accelerator as detailed in Sec.~\ref{sec:exp:standalone}, and we benchmark the performance under the NeRF-Synthetic Dataset~\cite{mildenhall2021nerf}, where the previous NeRF rendering accelerators~\cite{} report their performance., Tthe results areis summarized in Tab.~\ref{tab:benchmark_nerf_accelerator}. It should be noted that GBU achieves the highest rendering quality, thanks to the adoption of the advanced 3D Gaussian algorithm. Moreover, GBU also outperforms previous NeRF accelerators in terms of rendering speed while requiring less area and energy consumption, once again demonstrating the effectiveness of our proposed techniques.}

\noindent\textbf{Comparison with Standalone Accelerators.} This section benchmarks the proposed GBU against standalone 3D Gaussian and Neural Radience Field (NeRF) accelerators. It is important to note that GBU and standalone accelerators are not directly comparable, as the latter~\cite{lee2024gscore,rao2022icarus,li2022rt-nerf} provides end-to-end acceleration but typically specializes in only one type of scene (e.g., the static scene). In contrast, GBU accelerates only one rendering step and is compatible with a variety of AR/VR applications. For a fair comparison, we created a standalone version of GBU, dubbed GBU-Standalone, specifically for static scene rendering. GBU-Standalone is built by integrating GBU with dedicated hardware modules for \textit{Rendering Step \ding{182} and \ding{183}}. The implementation of these modules follows the design of GS-Core's Culling/Conversion/Sorting units~\cite{lee2024gscore} with the same setup in Sec.~\ref{sec:exp:setup} for evaluation.

\cready{As shown in Tab.~\ref{tab:hardware_standalone}, under the same target rendering speed in both the Tanks\&Temples dataset~\cite{Knapitsch2017} and the Deep Blending dataset~\cite{DeepBlending2018} used by GS-Core, GBU-Standalone demonstrates superior area and energy efficiency, primarily due to the proposed Tile Engine. Additionally, benchmarking against representative NeRF accelerators~\cite{rao2022icarus, li2022rt, li2023instant} on the NeRF-Synthetic dataset~\cite{mildenhall2021nerf} (Tab.~\ref{tab:benchmark_nerf_accelerator}) shows that GBU achieves the highest rendering quality, thanks to the advanced 3D Gaussian rendering algorithm, while also outperforming prior NeRF accelerators in rendering speed, area efficiency, and energy consumption, further validating the effectiveness of the proposed techniques.}

% \cready{We also benchmark GBU-Standalone with several representative NeRF accelerators under the NeRF-Synthetic dataset~\cite{mildenhall2021nerf}, which is commonly used by these NeRF accelerators. The results are summarized in Tab.~\ref{tab:benchmark_nerf_accelerator}. We observe that (1) GBU achieves the highest rendering quality, thanks to the adoption of the advanced 3D Gaussian rendering algorithm; and (2) GBU also outperforms previous NeRF accelerators in terms of rendering speed while requiring less area and energy consumption, once again demonstrating the effectiveness of the proposed techniques.}

\vspace{0.5em}
\noindent
\textbf{Limitations in extreme cases.} \cready{While GBU demonstrates strong performance across three widely used datasets~\cite{barron2021mipnerf,Li_2022_CVPR,alldieck2018video}, it may face challenges under certain extreme conditions:
(1) \textit{\underline{Distant camera poses.}} The efficiency of the IRSS dataflow relies on each Gaussian covering multiple pixels per row. However, when the camera is significantly farther from the scene, Gaussians may cover fewer pixels, reducing compute sharing. For instance, increasing the camera-to-scene distance by 4× in the static scene dataset~\cite{barron2021mipnerf} reduces GBU’s speedup over a vanilla GPU~\cite{JetsonOr73:online} from the original 10.8× to 4.7×. Future work could address this by adaptively merging Gaussians based on camera distance~\cite{hierarchicalgaussians24};
(2) \textit{\underline{Highly dynamic scenes.}} GBU primarily accelerates the \textit{Rendering Step \ding{184}}, but in highly dynamic scenes, other rendering steps may dominate computation. For example, multi-avatar settings~\cite{multi_avatar} may require substantial processing in the \textit{Rendering Step \ding{182}} for modeling the human bodies, limiting GBU’s overall speedup. A specialized accelerator for the \textit{Rendering Step \ding{182}} could improve efficiency in such scenarios.}

% From Fig.~\ref{fig:end-to-end-fps}, it can be observed that the proposed technique achieves the least speedup on the human avatar dataset~\cite{alldieck2018video}. This is because \textit{Rendering Stage \ding{182}} takes the highest percentage of latency on this dataset (see Fig.~\ref{fig:profiling_breakdown_2}) due to the complexity of modeling human body movement. For scenes requiring more detailed modeling of object movement and deformation, such as wrinkles on garments~\cite{rong2024gaussian} and deformations of human hair~\cite{zakharov2024gh}, we expect this rendering stage to become an even more severe bottleneck, potentially impeding real-time rendering. To overcome this future challenge, dedicated hardware and/or more efficient algorithms for modeling scene dynamics are needed.

% \vspace{2em}
\section{Related Works}

\noindent\textbf{Graphics Representations in 3D Reconstruction} Recently, NeRFs~\cite{mildenhall2021nerf} have demonstrated exceptional reconstruction quality. NeRFs employ implicit neural representations, parameterized by multi-layer perceptrons (MLPs), to model scenes. In the last year, Gaussian Splatting~\cite{kerbl3Dgaussians} has emerged as a novel 3D representation, striking the SOTA balance between real-time rendering and high reconstruction quality. This approach represents a scene as a collection of translucent 3D Gaussian kernels. During rendering, these 3D Gaussian kernels are projected as 2D Gaussian kernels onto a screen and then blended in screen space, alleviating the need for resource-intensive sampling in 3D. Given its effectiveness, Gaussian Splatting has been adapted for a variety of AR/VR applications, including video reconstruction\cite{luiten2023dynamic}, simultaneous localization and mapping (SLAM)\cite{yan2024gsslam}, 3D AI-generated content (AIGC)\cite{li2023gaussiandiffusion}, and virtual telepresence\cite{qian2023gaussianavatars}. The applications converge on a shared rendering pipeline that transforms these kernels into 2D images. Consequently, our GBU offers a unified solution for enhancing the performance of these AR/VR applications that are highly desirable for on-device deployment.

\noindent \textbf{Graphics Hardware} Researchers have developed specialized hardware accelerators dedicated to NeRFs~\cite{metavrain, li2022rt-nerf, li2023instant, lee2023neurex, gen_nerf, mubarik2023hardware}. These accelerators significantly outperform software rendering methods in both speed and energy efficiency. However, it is commonly agreed that the rendering processes for NeRFs and Gaussian Splatting are fundamentally different~\cite{kerbl3Dgaussians}; NeRFs require extensive sampling in 3D space, while Gaussian Splatting streamlines this process by directly rasterizing Gaussians onto a 2D screen. Consequently, accelerators and methodologies optimized for NeRFs are not directly transferable to Gaussian Splatting. This discrepancy underscores the need for a dedicated accelerator designed explicitly for Gaussian Splatting, ensuring real-time and high-fidelity rendering.

% The Graphics Processing Unit (GPU)~\cite{gpu} is the most commonly used hardware for graphics rendering. Frameworks such as OpenGL\cite{opengl} and Vulkan~\cite{vulkan} utilize the GPU's specialized hardware units, including rasterizers and texture filters, to enable real-time rendering. However, these units are primarily tailored for rendering textured meshes, which leads to suboptimal performance when rendering other graphics representations. In contrast, software rendering approaches~\cite{freepipe, laine2011high} use programmable shaders to simulate the functionalities of these specialized hardware units. This method becomes particularly useful when the rendering pipelines are incompatible with the specialized hardware units, as seen in the cases of Neural Radiance Fields (NeRFs) and Gaussian Splatting. However, empirical studies~\cite{laine2011high} have shown that even highly optimized software renderers can be 2 to 8 times slower than specialized hardware. As a result, novel geometric representations like NeRFs and Gaussian Splatting struggle to achieve real-time rendering on mobile GPUs without compromising quality, as shown in Sec.~\ref{sec:profiling}.

% To address this issue, r

% GPUs are the most widely adopted rendering hardware (a short review on GPU history).
% Software rendering
% NeRF Accelerators, specific to some rendering pipelines, not support gaussian.

\section{Conclusion}

Achieving real-time rendering speeds on edge devices remains a significant challenge due to the substantial computational demands associated with SOTA Gaussian-based rendering pipelines. In this work, we develop GBU, a hardware module specifically designed for edge systems to tackle these computational challenges. Our approach involves a comprehensive analysis of rendering pipelines in AR/VR applications to identify performance bottlenecks. Secondly, we develop a specialized dataflow that reduces the computational cost. Thirdly, we co-design a dedicated hardware module that seamlessly integrates into existing GPU architectures, improving data locality and leveraging a Gaussian Reuse Cache to optimize the rendering process. Extensive evaluations across various AR/VR applications demonstrate that the GBU not only addresses the primary latency bottlenecks but also supports a wide range of applications while maintaining SOTA rendering quality. These results confirm the effectiveness of our hardware-software co-design approach in bridging the performance gap on edge devices, paving the way for more immersive and responsive AR/VR experiences.

\vspace{0.2em}
\section*{ACKNOWLEDGMENTS}
\vspace{0.2em}
\cready{This work was supported by the National Science Foundation (NSF) Computing and Communication Foundations (CCF) program (Award IDs: 2400511 and 2312758), and CoCoSys, one of the seven centers in JUMP 2.0, a Semiconductor Research Corporation (SRC) program sponsored by DARPA.}

\bibliographystyle{IEEEtranS}
\bibliography{refs}

% Generated by IEEEtranS.bst, version: 1.13 (2008/09/30)
\begin{thebibliography}{10}
\providecommand{\url}[1]{#1}
\csname url@samestyle\endcsname
\providecommand{\newblock}{\relax}
\providecommand{\bibinfo}[2]{#2}
\providecommand{\BIBentrySTDinterwordspacing}{\spaceskip=0pt\relax}
\providecommand{\BIBentryALTinterwordstretchfactor}{4}
\providecommand{\BIBentryALTinterwordspacing}{\spaceskip=\fontdimen2\font plus
\BIBentryALTinterwordstretchfactor\fontdimen3\font minus \fontdimen4\font\relax}
\providecommand{\BIBforeignlanguage}[2]{{%
\expandafter\ifx\csname l@#1\endcsname\relax
\typeout{** WARNING: IEEEtranS.bst: No hyphenation pattern has been}%
\typeout{** loaded for the language `#1'. Using the pattern for}%
\typeout{** the default language instead.}%
\else
\language=\csname l@#1\endcsname
\fi
#2}}
\providecommand{\BIBdecl}{\relax}
\BIBdecl

\bibitem{AppleVis7}
``Apple vision pro - apple,'' \url{https://www.apple.com/apple-vision-pro/}, (Accessed on 04/09/2024).

\bibitem{JetsonOr73:online}
``Jetson orin for next-gen robotics | nvidia,'' \url{https://www.nvidia.com/en-us/autonomous-machines/embedded-systems/jetson-orin/}, (Accessed on 04/02/2024).

\bibitem{MetaQues16}
``Meta quest 3: New mixed reality vr headset - shop now | meta store | meta store,'' \url{https://www.meta.com/quest/quest-3/}, (Accessed on 04/09/2024).

\bibitem{NsightSy30:online}
``Nsight systems — nsight-systems 2024.2 documentation,'' \url{https://docs.nvidia.com/nsight-systems/index.html}, (Accessed on 04/16/2024).

\bibitem{alldieck2018video}
T.~Alldieck, M.~Magnor, W.~Xu, C.~Theobalt, and G.~Pons-Moll, ``Video based reconstruction of 3d people models,'' in \emph{{IEEE}/{CVF} Conference on Computer Vision and Pattern Recognition ({CVPR})}, Jun 2018, pp. 8387--8397, {CVPR} Spotlight Paper.

\bibitem{attal2023hyperreel}
B.~Attal, J.-B. Huang, C.~Richardt, M.~Zollhoefer, J.~Kopf, M.~O'Toole, and C.~Kim, ``{HyperReel}: {H}igh-fidelity {6-DoF} video with ray-conditioned sampling,'' \emph{arXiv preprint arXiv:2301.02238}, 2023.

\bibitem{barron2021mipnerf}
J.~T. Barron, B.~Mildenhall, M.~Tancik, P.~Hedman, R.~Martin-Brualla, and P.~P. Srinivasan, ``Mip-nerf: A multiscale representation for anti-aliasing neural radiance fields,'' 2021.

\bibitem{Chen2022ECCV}
A.~Chen, Z.~Xu, A.~Geiger, J.~Yu, and H.~Su, ``Tensorf: Tensorial radiance fields,'' in \emph{European Conference on Computer Vision (ECCV)}, 2022.

\bibitem{chen2024survey}
G.~Chen and W.~Wang, ``A survey on 3d gaussian splatting,'' \emph{arXiv preprint arXiv:2401.03890}, 2024.

\bibitem{chen2021animatable}
J.~Chen, Y.~Zhang, D.~Kang, X.~Zhe, L.~Bao, X.~Jia, and H.~Lu, ``Animatable neural radiance fields from monocular rgb videos,'' 2021.

\bibitem{chen2023gaussianeditor}
Y.~Chen, Z.~Chen, C.~Zhang, F.~Wang, X.~Yang, Y.~Wang, Z.~Cai, L.~Yang, H.~Liu, and G.~Lin, ``Gaussianeditor: Swift and controllable 3d editing with gaussian splatting,'' \emph{arXiv preprint arXiv:2311.14521}, 2023.

\bibitem{yu2022plenoxels}
S.~Fridovich-Keil, A.~Yu, M.~Tancik, Q.~Chen, B.~Recht, and A.~Kanazawa, ``Plenoxels: Radiance fields without neural networks,'' in \emph{CVPR}, 2022.

\bibitem{gen_nerf}
\BIBentryALTinterwordspacing
Y.~Fu, Z.~Ye, J.~Yuan, S.~Zhang, S.~Li, H.~You, and Y.~Lin, ``Gen-nerf: Efficient and generalizable neural radiance fields via algorithm-hardware co-design,'' in \emph{Proceedings of the 50th Annual International Symposium on Computer Architecture}, ser. ISCA '23.\hskip 1em plus 0.5em minus 0.4em\relax New York, NY, USA: Association for Computing Machinery, 2023. [Online]. Available: \url{https://doi.org/10.1145/3579371.3589109}
\BIBentrySTDinterwordspacing

\bibitem{greub2012linear}
W.~H. Greub, \emph{Linear algebra}.\hskip 1em plus 0.5em minus 0.4em\relax Springer Science \& Business Media, 2012, vol.~23.

\bibitem{metavrain}
D.~Han, J.~Ryu, S.~Kim, S.~Kim, J.~Park, and H.-J. Yoo, ``Metavrain: A mobile neural 3-d rendering processor with bundle-frame-familiarity-based nerf acceleration and hybrid dnn computing,'' \emph{IEEE Journal of Solid-State Circuits}, vol.~59, no.~1, pp. 65--78, 2024.

\bibitem{DeepBlending2018}
P.~Hedman, J.~Philip, T.~Price, J.-M. Frahm, G.~Drettakis, and G.~Brostow, ``Deep blending for free-viewpoint image-based rendering,'' vol.~37, no.~6, pp. 257:1--257:15, 2018.

\bibitem{hu2023gauhuman}
S.~Hu and Z.~Liu, ``Gauhuman: Articulated gaussian splatting from monocular human videos,'' \emph{arXiv preprint arXiv:2312.02973}, 2023.

\bibitem{huang2023sc}
Y.-H. Huang, Y.-T. Sun, Z.~Yang, X.~Lyu, Y.-P. Cao, and X.~Qi, ``Sc-gs: Sparse-controlled gaussian splatting for editable dynamic scenes,'' \emph{arXiv preprint arXiv:2312.14937}, 2023.

\bibitem{jiang2022instantavatar}
T.~Jiang, X.~Chen, J.~Song, and O.~Hilliges, ``Instantavatar: Learning avatars from monocular video in 60 seconds,'' \emph{arXiv}, 2022.

\bibitem{kerbl3Dgaussians}
\BIBentryALTinterwordspacing
B.~Kerbl, G.~Kopanas, T.~Leimk{\"u}hler, and G.~Drettakis, ``3d gaussian splatting for real-time radiance field rendering,'' \emph{ACM Transactions on Graphics}, vol.~42, no.~4, July 2023. [Online]. Available: \url{https://repo-sam.inria.fr/fungraph/3d-gaussian-splatting/}
\BIBentrySTDinterwordspacing

\bibitem{hierarchicalgaussians24}
\BIBentryALTinterwordspacing
B.~Kerbl, A.~Meuleman, G.~Kopanas, M.~Wimmer, A.~Lanvin, and G.~Drettakis, ``A hierarchical 3d gaussian representation for real-time rendering of very large datasets,'' \emph{ACM Transactions on Graphics}, vol.~43, no.~4, July 2024. [Online]. Available: \url{https://repo-sam.inria.fr/fungraph/hierarchical-3d-gaussians/}
\BIBentrySTDinterwordspacing

\bibitem{khairy2020accel}
M.~Khairy, Z.~Shen, T.~M. Aamodt, and T.~G. Rogers, ``Accel-sim: An extensible simulation framework for validated gpu modeling,'' in \emph{2020 ACM/IEEE 47th Annual International Symposium on Computer Architecture (ISCA)}.\hskip 1em plus 0.5em minus 0.4em\relax IEEE, 2020, pp. 473--486.

\bibitem{Knapitsch2017}
A.~Knapitsch, J.~Park, Q.-Y. Zhou, and V.~Koltun, ``Tanks and temples: Benchmarking large-scale scene reconstruction,'' \emph{ACM Transactions on Graphics}, vol.~36, no.~4, 2017.

\bibitem{lee2023neurex}
J.~Lee, K.~Choi, J.~Lee, S.~Lee, J.~Whangbo, and J.~Sim, ``Neurex: A case for neural rendering acceleration,'' in \emph{Proceedings of the 50th Annual International Symposium on Computer Architecture}, 2023, pp. 1--13.

\bibitem{lee2024gscore}
J.~Lee, S.~Lee, J.~Lee, J.~Park, and J.~Sim, ``Gscore: Efficient radiance field rendering via architectural support for 3d gaussian splatting,'' in \emph{Proceedings of the 29th ACM International Conference on Architectural Support for Programming Languages and Operating Systems, Volume 3}, 2024, pp. 497--511.

\bibitem{lei2023gart}
J.~Lei, Y.~Wang, G.~Pavlakos, L.~Liu, and K.~Daniilidis, ``Gart: Gaussian articulated template models,'' 2023.

\bibitem{li2022rt}
C.~Li, S.~Li, Y.~Zhao, W.~Zhu, and Y.~Lin, ``Rt-nerf: Real-time on-device neural radiance fields towards immersive ar/vr rendering,'' in \emph{Proceedings of the 41st IEEE/ACM International Conference on Computer-Aided Design}, 2022, pp. 1--9.

\bibitem{li2022rt-nerf}
C.~Li, S.~Li, Y.~Zhao, W.~Zhu, and Y.~Lin, ``Rt-nerf: Real-time on-device neural radiance fields towards immersive ar/vr rendering,'' \emph{IEEE/ACM International Conference on Computer-Aided Design (ICCAD 2022)}, 2022.

\bibitem{li2023human101}
M.~Li, J.~Tao, Z.~Yang, and Y.~Yang, ``Human101: Training 100+fps human gaussians in 100s from 1 view,'' 2023.

\bibitem{li2023instant}
S.~Li, C.~Li, W.~Zhu, B.~Yu, Y.~Zhao, C.~Wan, H.~You, H.~Shi, and Y.~Lin, ``Instant-3d: Instant neural radiance field training towards on-device ar/vr 3d reconstruction,'' in \emph{Proceedings of the 50th Annual International Symposium on Computer Architecture}, 2023, pp. 1--13.

\bibitem{li2022neural}
T.~Li, M.~Slavcheva, M.~Zollhoefer, S.~Green, C.~Lassner, C.~Kim, T.~Schmidt, S.~Lovegrove, M.~Goesele, R.~Newcombe, and Z.~Lv, ``Neural 3d video synthesis from multi-view video,'' 2022.

\bibitem{Li_2022_CVPR}
T.~Li, M.~Slavcheva, M.~Zollh\"ofer, S.~Green, C.~Lassner, C.~Kim, T.~Schmidt, S.~Lovegrove, M.~Goesele, R.~Newcombe, and Z.~Lv, ``Neural 3d video synthesis from multi-view video,'' in \emph{Proceedings of the IEEE/CVF Conference on Computer Vision and Pattern Recognition (CVPR)}, June 2022, pp. 5521--5531.

\bibitem{li2023gaussiandiffusion}
X.~Li, H.~Wang, and K.-K. Tseng, ``Gaussiandiffusion: 3d gaussian splatting for denoising diffusion probabilistic models with structured noise,'' 2023.

\bibitem{multi_avatar}
\BIBentryALTinterwordspacing
Y.~Liu, X.~Huang, M.~Qin, Q.~Lin, and H.~Wang, ``Animatable 3d gaussian: Fast and high-quality reconstruction of multiple human avatars,'' in \emph{Proceedings of the 32nd ACM International Conference on Multimedia}, ser. MM '24.\hskip 1em plus 0.5em minus 0.4em\relax New York, NY, USA: Association for Computing Machinery, 2024, p. 1120–1129. [Online]. Available: \url{https://doi.org/10.1145/3664647.3680674}
\BIBentrySTDinterwordspacing

\bibitem{lombardi2018deep}
S.~Lombardi, J.~Saragih, T.~Simon, and Y.~Sheikh, ``Deep appearance models for face rendering,'' \emph{ACM Transactions on Graphics (ToG)}, vol.~37, no.~4, pp. 1--13, 2018.

\bibitem{luiten2023dynamic}
J.~Luiten, G.~Kopanas, B.~Leibe, and D.~Ramanan, ``Dynamic 3d gaussians: Tracking by persistent dynamic view synthesis,'' in \emph{3DV}, 2024.

\bibitem{Ma_2021_CVPR}
S.~Ma, T.~Simon, J.~Saragih, D.~Wang, Y.~Li, F.~De~la Torre, and Y.~Sheikh, ``Pixel codec avatars,'' in \emph{Proceedings of the IEEE/CVF Conference on Computer Vision and Pattern Recognition (CVPR)}, June 2021, pp. 64--73.

\bibitem{mildenhall2021nerf}
B.~Mildenhall, P.~P. Srinivasan, M.~Tancik, J.~T. Barron, R.~Ramamoorthi, and R.~Ng, ``Nerf: Representing scenes as neural radiance fields for view synthesis,'' \emph{Communications of the ACM}, vol.~65, no.~1, pp. 99--106, 2021.

\bibitem{mubarik2023hardware}
M.~H. Mubarik, R.~Kanungo, T.~Zirr, and R.~Kumar, ``Hardware acceleration of neural graphics,'' in \emph{Proceedings of the 50th Annual International Symposium on Computer Architecture}, 2023, pp. 1--12.

\bibitem{mueller2022instant}
\BIBentryALTinterwordspacing
T.~M\"uller, A.~Evans, C.~Schied, and A.~Keller, ``Instant neural graphics primitives with a multiresolution hash encoding,'' \emph{ACM Trans. Graph.}, vol.~41, no.~4, pp. 102:1--102:15, Jul. 2022. [Online]. Available: \url{https://doi.org/10.1145/3528223.3530127}
\BIBentrySTDinterwordspacing

\bibitem{holoportation}
\BIBentryALTinterwordspacing
S.~Orts-Escolano, C.~Rhemann, S.~Fanello, W.~Chang, A.~Kowdle, Y.~Degtyarev, D.~Kim, P.~L. Davidson, S.~Khamis, M.~Dou, V.~Tankovich, C.~Loop, Q.~Cai, P.~A. Chou, S.~Mennicken, J.~Valentin, V.~Pradeep, S.~Wang, S.~B. Kang, P.~Kohli, Y.~Lutchyn, C.~Keskin, and S.~Izadi, ``Holoportation: Virtual 3d teleportation in real-time,'' in \emph{UIST 2016}, March 2016. [Online]. Available: \url{https://www.microsoft.com/en-us/research/publication/holoportation-virtual-3d-teleportation-in-real-time/}
\BIBentrySTDinterwordspacing

\bibitem{Pumarola_2021_CVPR}
A.~Pumarola, E.~Corona, G.~Pons-Moll, and F.~Moreno-Noguer, ``D-nerf: Neural radiance fields for dynamic scenes,'' in \emph{Proceedings of the IEEE/CVF Conference on Computer Vision and Pattern Recognition (CVPR)}, June 2021, pp. 10\,318--10\,327.

\bibitem{qian2023gaussianavatars}
S.~Qian, T.~Kirschstein, L.~Schoneveld, D.~Davoli, S.~Giebenhain, and M.~Nie\ss{}ner, ``Gaussianavatars: Photorealistic head avatars with rigged 3d gaussians,'' \emph{arXiv preprint arXiv:2312.02069}, 2023.

\bibitem{qin2023langsplat}
M.~Qin, W.~Li, J.~Zhou, H.~Wang, and H.~Pfister, ``Langsplat: 3d language gaussian splatting,'' \emph{arXiv preprint arXiv:2312.16084}, 2023.

\bibitem{rao2022icarus}
C.~Rao, H.~Yu, H.~Wan, J.~Zhou, Y.~Zheng, M.~Wu, Y.~Ma, A.~Chen, B.~Yuan, P.~Zhou \emph{et~al.}, ``Icarus: A specialized architecture for neural radiance fields rendering,'' \emph{ACM Transactions on Graphics (TOG)}, vol.~41, no.~6, pp. 1--14, 2022.

\bibitem{SplattingAvatar}
Z.~Shao, Z.~Wang, Z.~Li, D.~Wang, X.~Lin, Y.~Zhang, M.~Fan, and Z.~Wang, ``{SplattingAvatar: Realistic Real-Time Human Avatars with Mesh-Embedded Gaussian Splatting},'' in \emph{Computer Vision and Pattern Recognition (CVPR)}, 2024.

\bibitem{tang2023dreamgaussian}
J.~Tang, J.~Ren, H.~Zhou, Z.~Liu, and G.~Zeng, ``Dreamgaussian: Generative gaussian splatting for efficient 3d content creation,'' \emph{arXiv preprint arXiv:2309.16653}, 2023.

\bibitem{wang2022mixed}
F.~Wang, S.~Tan, X.~Li, Z.~Tian, and H.~Liu, ``Mixed neural voxels for fast multi-view video synthesis,'' \emph{arXiv preprint arXiv:2212.00190}, 2022.

\bibitem{wu20234dgaussians}
G.~Wu, T.~Yi, J.~Fang, L.~Xie, X.~Zhang, W.~Wei, W.~Liu, Q.~Tian, and W.~Xinggang, ``4d gaussian splatting for real-time dynamic scene rendering,'' \emph{arXiv preprint arXiv:2310.08528}, 2023.

\bibitem{yan2024gsslam}
C.~Yan, D.~Qu, D.~Wang, D.~Xu, Z.~Wang, B.~Zhao, and X.~Li, ``Gs-slam: Dense visual slam with 3d gaussian splatting,'' 2024.

\bibitem{yang2023gs4d}
Z.~Yang, H.~Yang, Z.~Pan, and L.~Zhang, ``Real-time photorealistic dynamic scene representation and rendering with 4d gaussian splatting,'' 2024.

\bibitem{zhang2024gaussiancube}
B.~Zhang, Y.~Cheng, J.~Yang, C.~Wang, F.~Zhao, Y.~Tang, D.~Chen, and B.~Guo, ``Gaussiancube: Structuring gaussian splatting using optimal transport for 3d generative modeling,'' \emph{arXiv preprint arXiv:2403.19655}, 2024.

\bibitem{zhou2024drivinggaussian}
X.~Zhou, Z.~Lin, X.~Shan, Y.~Wang, D.~Sun, and M.-H. Yang, ``Drivinggaussian: Composite gaussian splatting for surrounding dynamic autonomous driving scenes,'' 2024.

\bibitem{framerate}
D.~J. Zielinski, H.~M. Rao, M.~A. Sommer, and R.~Kopper, ``Exploring the effects of image persistence in low frame rate virtual environments,'' in \emph{2015 IEEE Virtual Reality (VR)}, 2015, pp. 19--26.

\bibitem{zielonka2023drivable}
W.~Zielonka, T.~Bagautdinov, S.~Saito, M.~Zollh{\"o}fer, J.~Thies, and J.~Romero, ``Drivable 3d gaussian avatars,'' \emph{arXiv preprint arXiv:2311.08581}, 2023.

\bibitem{zwicker2001ewa}
M.~Zwicker, H.~Pfister, J.~Van~Baar, and M.~Gross, ``Ewa volume splatting,'' in \emph{Proceedings Visualization, 2001. VIS'01.}\hskip 1em plus 0.5em minus 0.4em\relax IEEE, 2001, pp. 29--538.

\end{thebibliography}
%%%%%%%%%%%%%%%%%%%%%%%%%%%%%%%%%%%%

\end{document}